\documentclass[sigconf]{acmart}
\usepackage{enumitem}
\usepackage{multirow}
\usepackage{subfigure}
\usepackage{bbm}
\usepackage{abraces}
\usepackage[ruled,vlined]{algorithm2e}
\newtheorem{observation}{Observation}

\newtheorem{theorem}{Theorem}
\newtheorem{assumption}{Assumption}

\AtBeginDocument{%
  \providecommand\BibTeX{{%
    \normalfont B\kern-0.5em{\scshape i\kern-0.25em b}\kern-0.8em\TeX}}}

\copyrightyear{2022}
\acmYear{2022}
\setcopyright{acmcopyright}\acmConference[CIKM '22]{Proceedings of the 31st ACM International Conference on Information and Knowledge Management}{October 17--21, 2022}{Atlanta, GA, USA}
\acmBooktitle{Proceedings of the 31st ACM International Conference on Information and Knowledge Management (CIKM '22), October 17--21, 2022, Atlanta, GA, USA}
\acmPrice{15.00}
\acmDOI{10.1145/3511808.3557462}
\acmISBN{978-1-4503-9236-5/22/10}

\begin{document}

\title{SVD-GCN: A Simplified Graph Convolution Paradigm for Recommendation}


\author{Shaowen Peng}
\email{swpeng95@gmail.com}
\affiliation{%
  \institution{Kyoto University}
  \city{Kyoto}
  \country{Japan}
}

\author{Kazunari Sugiyama}
\email{kaz.sugiyama@i.kyoto-u.ac.jp}
\affiliation{%
  \institution{Kyoto University}
  \city{Kyoto}
  \country{Japan}
}

\author{Tsunenori Mine}
\email{mine@m.ait.kyushu-u.ac.jp}
\affiliation{%
  \institution{Kyushu University}
  \city{Fukuoka}
  \country{Japan}
}

\begin{abstract}
With the tremendous success of Graph Convolutional Networks (GCNs), they have been widely applied to recommender systems and have shown promising performance. However, most GCN-based methods rigorously stick to a common GCN learning paradigm and suffer from two limitations: (1) the 
limited scalability due to the high computational cost and slow training convergence; (2) the notorious over-smoothing issue which reduces performance as stacking graph convolution layers. We argue that the above limitations are due to the lack of a deep understanding of GCN-based methods. To this end, we first investigate what design makes GCN effective for recommendation. By simplifying LightGCN, we show the close connection between GCN-based and low-rank methods such as Singular Value Decomposition (SVD) and Matrix Factorization (MF), where stacking graph convolution layers is to learn a low-rank representation by emphasizing (suppressing) components with larger (smaller) singular values. Based on this observation, we replace the core design of GCN-based methods with a flexible truncated SVD and propose a simplified GCN learning paradigm dubbed SVD-GCN, which only exploits $K$-largest singular vectors for recommendation. To alleviate the over-smoothing issue, we propose a renormalization trick to adjust the singular value gap, resulting in significant improvement. Extensive experiments on three real-world datasets show that our proposed SVD-GCN not only significantly outperforms state-of-the-arts but also achieves over 100x and 10x speedups over LightGCN and MF, respectively.        
  
\end{abstract}



\maketitle

\section{Introduction}
With rapid development of the Internet and web services, recommender systems have been playing an important role in people's daily life. As a fundamental task for recommendation, Collaborative Filtering (CF) focuses on digging out the user preference from past user-item interactions, and has received much attention for decades. One of the most widely used CF methods, low-rank matrix factorization (MF) \cite{koren2009matrix}, characterizes user/item as latent vectors in an embedding space and estimates ratings as the cosine similarity between user and item latent vectors. To overcome the drawback of MF that a linear function is inefficient to capture complex user behaviour, subsequent works incorporate side information (e.g., user reviews, image data, temporal information, etc.) \cite{chen2017attentive,jiang2012social,feng2017poi2vec} and exploit advanced algorithms \cite{he2017neural,sun2019bert4rec,wang2017irgan} to infer user preference.\par

However, traditional CF methods heavily rely upon the quality of interactions as they can only learn the direct user-item relations. Therefore, they always show poor performance due to the common data sparsity issue in practice. Recently, Graph Convolutional Networks (GCNs) \cite{kipf2017semi} have shown great potential in various fields including social network analysis \cite{fan2019graph,wu2019neural} and recommender systems \cite{berg2017graph,ying2018graph}. Much research effort has been devoted to adapt GCNs for recommendation, such as augmenting GCNs with other advanced algorithms \cite{sun2021hgcf,wu2020self,ji2020dual}, simplifying GCNs to improve training efficiency and model effectiveness \cite{he2020lightgcn,mao2021ultragcn,chen2020revisiting}, and so on. By representing user-item interactions as a bipartite graph, the core idea of GCNs is to repeatedly propagate user and item embeddings on the graph to aggregate higher-order collaborative signals, thereby learning high quality embeddings even with limited interactions. Despite its effectiveness, most existing GCN-based methods suffer from the following limitations:
\begin{itemize}[leftmargin=10pt]

\item The core step of GCNs is implemented by repeatedly multiplying by an adjacency matrix, resulting in high computational cost and poor scalability.  

\item As shown in many works \cite{li2018deeper,zhao2020pairnorm}, stacking graph convolution layers tends to cause the overs-smoothing issue, resulting in similar user/item representations and reducing the recommendation accuracy. As a result, most existing GCN-based CF methods remain shallow (two, three layers at most). 

\item Unlike traditional CF methods, user/item representations are contributed from tremendous higher-order neighborhood, making the model difficult to train. Some GCN-based CF methods such as LightGCN requires about 800 epochs to reach the best accuracy, which further increases the training cost.   

\end{itemize}
We argue that the above limitations are due to the lack of a deep understanding of GCNs. Thus, in this work, we aim to figure out: what is the core design making GCNs effective for recommendation? Based on our answer to this question, we propose a scalable and simple GCN learning paradigm without above limitations.\par

To this end, we first dissect LightGCN, a linear GCN-based CF method which only exploits neighborhood aggregation and removes other designs. By simplifying LightGCN, we show that it is closely related to low-rank CF methods such as Singular Value Decomposition (SVD) and low-rank Matrix Factorization (MF), where stacking graph convolution layers is to learn a low-rank representation by emphasizing (suppressing) the components with larger (smaller) singular values. With empirical analysis, we further show that only a very few components corresponding to $K$-largest singular values contribute to recommendation performance, whereas most information (over 95\% on the tested data) are noisy and can be removed. Based on the above analysis, we replace the core component of GCNs (i.e., neighborhood aggregation) with a flexible truncated SVD and propose a simplified GCN learning paradigm dubbed SVD-GCN. Specifically, SVD-GCN only requires a very few ($K$-largest) singular values (vectors) and model parameters (less than 1\% of MF's on the tested data) for prediction. To alleviate the over-smoothing issue, we propose a renormalization trick to adjust the singular value gap, making important features of interactions well preserved, thereby resulting in significant improvement. Furthermore, to make the best of interactions, we augment SVD-GCN with user-user and item-item relations, leading to further improvement. Since the superiority of GCNs over traditional CF methods lies in the ability to augment interactions with higher-order collaborative signals, we only use 20\% of the interactions for training to evaluate the robustness and effectiveness of GCN designs. The main contributions of this work are summarized as follows:

\begin{itemize}[leftmargin=10pt]

\item By showing the connection between GCN-based and low-rank CF methods, we provide deep insight into GCN-based CF methods, that they contribute to recommendation in the same way as low-rank methods.

\item Distinct from the GCN learning paradigm that most GCN-based methods rigorously sticking to, we propose a simplified formulation of GCNs dubbed SVD-GCN, which only exploits $K$-largest singular values and vectors and is equipped with a lighter structure than MF.

\item To tackle the over-smoothing issue, we propose a renormalization trick to adjust the singular value gap to assure that important features from interactions are well preserved, leading to significant improvement.

\item Extensive experiments on three datasets show that our proposed SVD-GCN outperforms state-of-the-art with higher training efficiency and less running time. 

\end{itemize}

\section{preliminaries}
\subsection{GCN learning paradigm for CF}
We summarize a common GCN learning paradigm for CF. Given the user set $\mathcal{U}$, item set $\mathcal{I}$ and an interaction matrix $\mathbf{R}\in \{0, 1\}^{\left | \mathcal{U} \right | \times \left | \mathcal{I} \right |}$, we define a bipartite graph $\mathcal{G}=(\mathcal{V}, \mathcal{E})$, where the node set $\mathcal{V}=\mathcal{U} \cup \mathcal{I}$ contains all users and items, the edge set $\mathcal{E}=\mathbf{R}^+$ is represented by observed interactions, where $\mathbf{R}^+=\{r_{ui}=1|u\in\mathcal{U}, i\in\mathcal{I}\}$. Each user/item is considered as a node on the graph and parameterized as an embedding vector $\mathbf{e}_u/\mathbf{e}_i\in\mathbb{R}^d$. The core idea of GCNs is to update user and item embeddings by propagating them on the graph. The adjacency relations are represented as:
\begin{equation}
\mathbf{A}=\begin{bmatrix}
 \mathbf{0}& \mathbf{R}\\ 
\mathbf{R}^T &\mathbf{0} 
\end{bmatrix}.
\label{adjacency}
\end{equation}
The updating rule of GCNs is formulated as follows:
\begin{equation}
\mathbf{H}^{(l+1)}=\sigma\left( \mathbf{\tilde{A}} \mathbf{H}^{(l)} \mathbf{W}^{(l+1)} \right),
\end{equation}  
where $\mathbf{\tilde{A}}=\mathbf{D}^{\mbox{-}\frac{1}{2}}\mathbf{A}\mathbf{D}^{\mbox{-}\frac{1}{2}}$ is a symmetric normalized adjacency matrix, $\mathbf{D}$ is a diagonal node degree matrix. The initial state is $\mathbf{H}^{(0)}=\mathbf{E}$, where $\mathbf{E}\in\mathbb{R}^{(\left|\mathcal{U}\right|+\left|\mathcal{I}\right|)\times d}$ contains users' and items' embedding vectors. Recent works \cite{he2020lightgcn,chen2020revisiting} show the non-linear activation function $\sigma(\cdot)$ and feature transformations $\mathbf{W}^{(l+1)}$ are redundant for CF , the above updating rule can be simplified as follows:
\begin{equation}
\mathbf{H}^{(l)}=\mathbf{\tilde{A}}^l \mathbf{E}.
\end{equation}  
The final embeddings are generated by accumulating the embeddings at each layer through a pooling function:
\begin{equation}
\mathbf{O}=\mathbf{pooling}\left(\mathbf{H}^{(l)}|l=\{0,1,\cdots,L\}\right).
\label{pooling}
\end{equation}
Finally, an interaction is estimated as the inner product between a user's and an item's final embedding:
\begin{equation}
\hat{r}_{ui}=\mathbf{o}_u^T\mathbf{o}_i.
\end{equation} 

\subsection{Low-Rank Methods}
Low rank representation plays a fundamental role in modern recommender systems \cite{jin2021towards}. The core idea of low-rank methods is inspired by Singular Value Decomposition (SVD):
\begin{equation}
\mathbf{R}=\mathbf{U}diag\left(s_k\right)\mathbf{V}^T\approx\sum_{k=1}^K s_k \mathbf{u}_k \mathbf{v}_k^T.
\label{svd}
\end{equation}
The interaction matrix can be decomposed to three matrices, where the column of [$\mathbf{U}$ and $\mathbf{V}$ (i.e., $\mathbf{u}_k$ and $\mathbf{v}_k$)] and $s_k$ are [left and right singular vectors] and singular value, respectively; $s_1>s_2>\cdots\geq0$; $diag(\cdot)$ is the diagonalization operation. Since the components with larger (smaller) singular values contribute more (less) to interactions, we can approximate $\mathbf{R}$ with only $K$-largest singular values. Alternatively, we can learn low-rank representations in a dynamical way through matrix factorization (MF) \cite{koren2009matrix}: 

\begin{equation}
{\rm min}\sum_{(u,i)\in\mathbf{R}^+} \left \| r_{ui}-\mathbf{e}_u^T\mathbf{e}_i \right \|^2_2+\lambda\left(\left \| \mathbf{e}_u\right\|^2_2 + \left \| \mathbf{e}_i\right\|^2_2 \right),
\label{MF}
\end{equation}
where $\lambda$ is the strength for regularization. Each user and item is represented as a trainable vector with dimension $d\leq {\rm min}(\left|\mathcal{U} \right|, \left|\mathcal{V} \right|)$. By optimizing the following objective function, the model is expected to learn important features from interactions (e.g., components corresponding to $d$-largest singular values).

\begin{figure*} \centering 
\subfigure[The accuracy of three variants on CiteULike evaluated by nDCG@10. ] {  
\includegraphics[width=0.45\columnwidth]{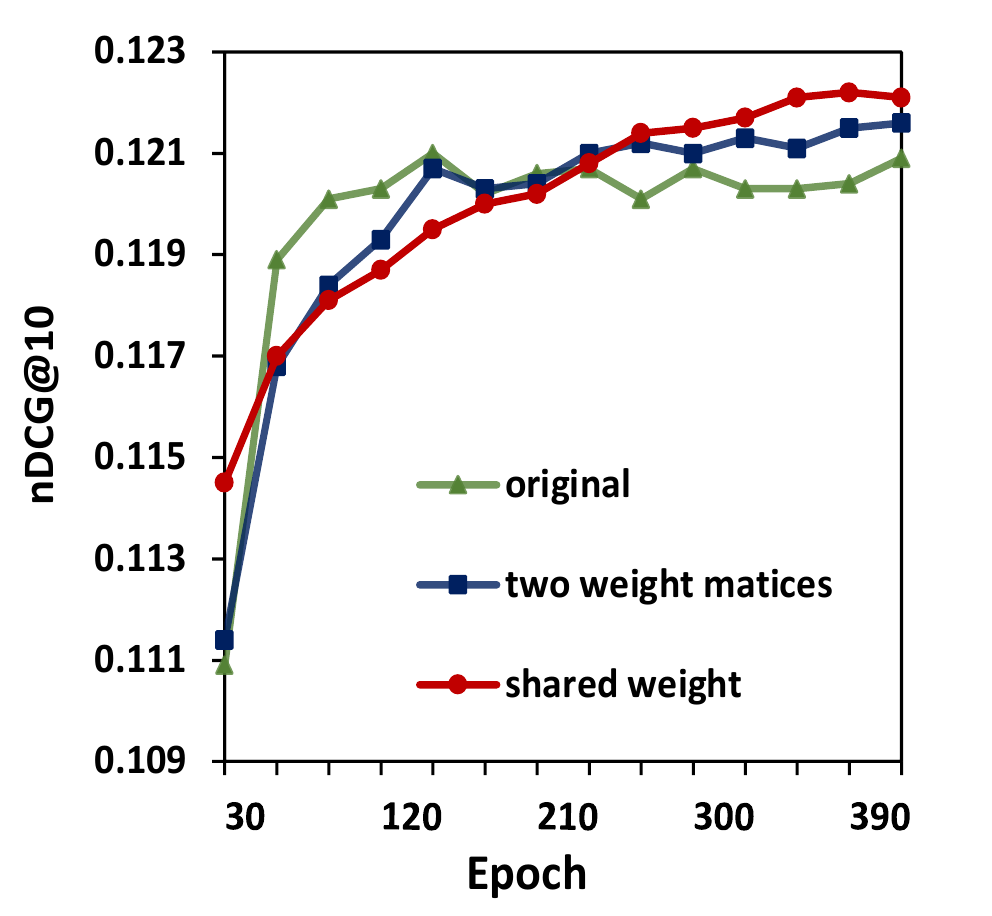} 
}
\hspace{0.05cm}
\subfigure[The accuracy of three variants on ML-100K. ] {  
\includegraphics[width=0.45\columnwidth]{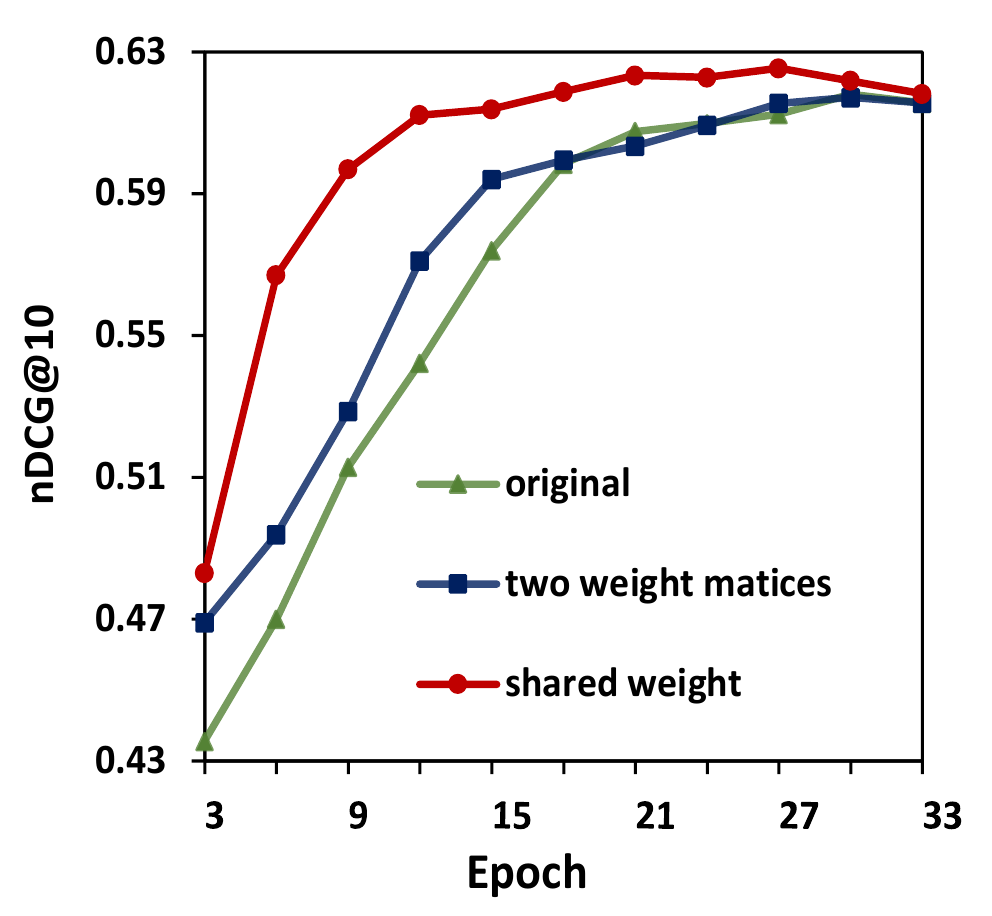} 
}
\subfigure[Comparison between SVD-LightGCN and SVD-LightGCN-T on CiteULike. ] {  
\includegraphics[width=0.45\columnwidth]{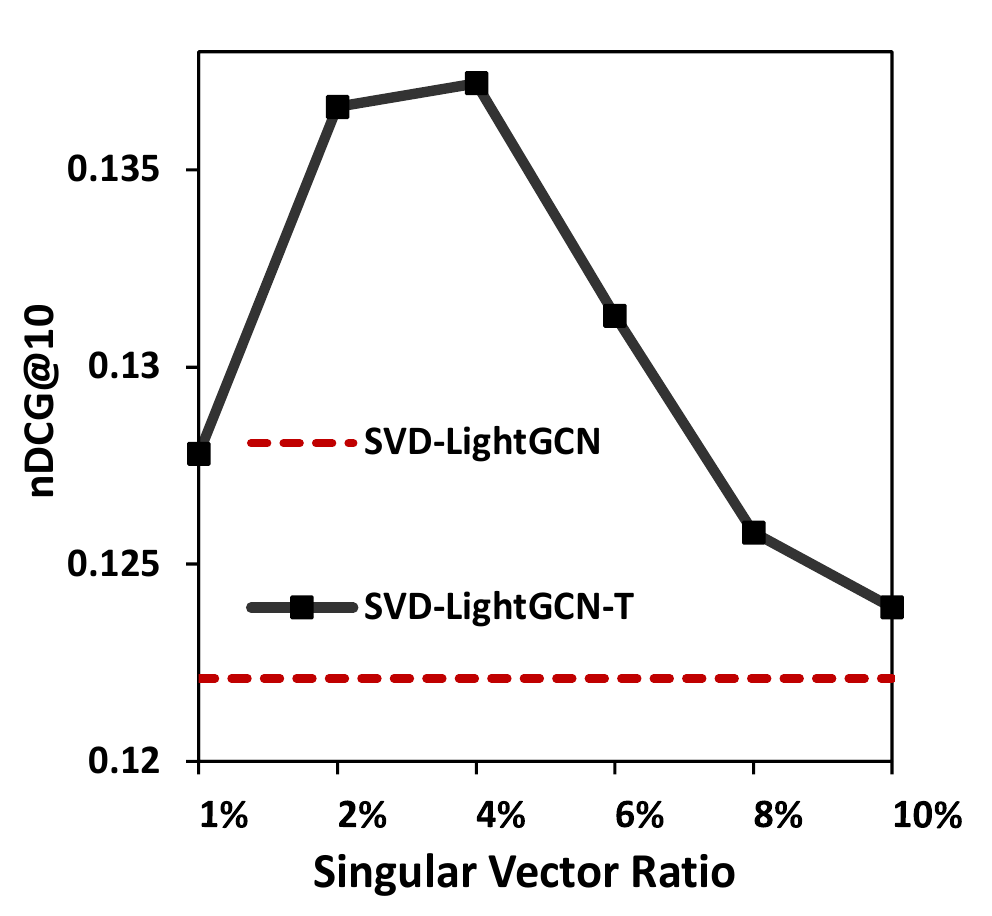} 
}
\hspace{0.05cm}
\subfigure[Comparison between SVD-LightGCN and SVD-LightGCN-T on ML-100K. ] {  
\includegraphics[width=0.45\columnwidth]{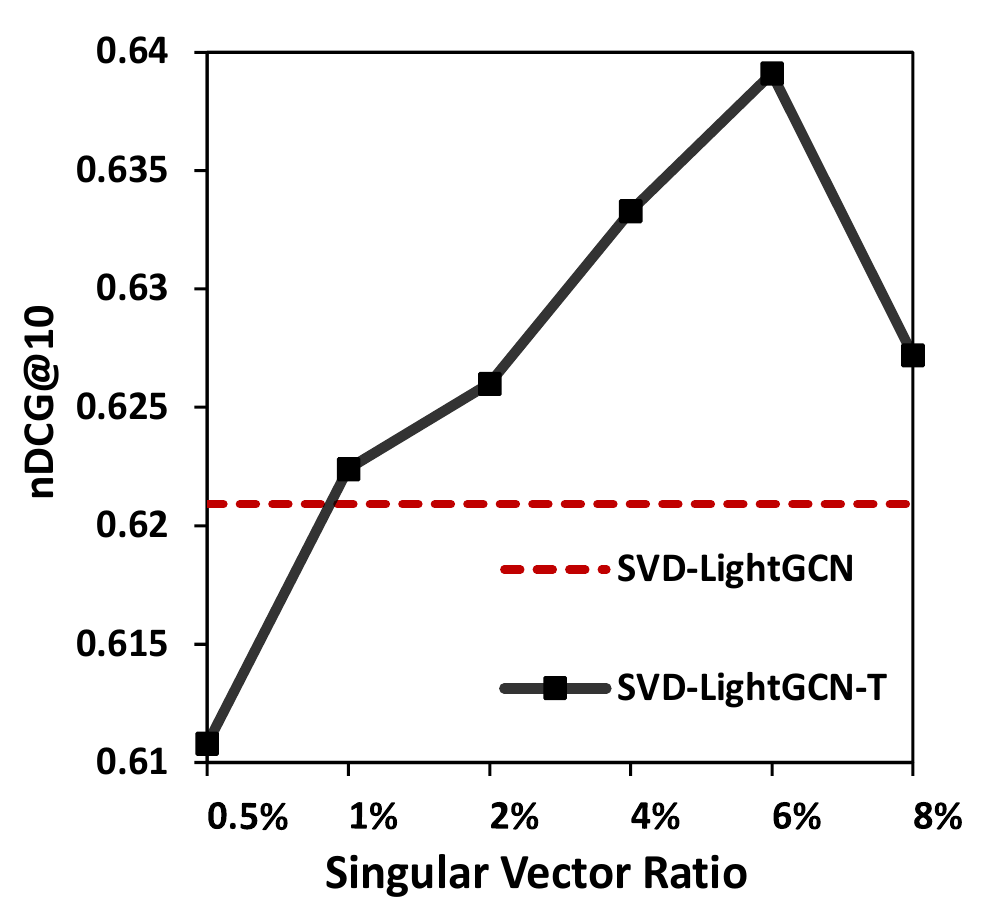} 
}
\vspace{-0.2cm}
\caption{Some empirical results on two datasets (CiteULike and ML-100K).}
\label{analysis}
\end{figure*}

\section{methodology}
\subsection{Connections Between GCNs and SVD}
As activation functions and feature transformations have been shown ineffective for CF \cite{he2020lightgcn}, we focus on LightGCN whose final embeddings are generated as follows:
\begin{equation}
\mathbf{O}=\sum_{l=0}^L \frac{\mathbf{H}^{(l)}}{L+1}=\left(\sum_{l=0}^L\frac{\mathbf{\tilde{A}}^l}{L+1} \right) \mathbf{E},
\label{lightgcn}
\end{equation}
where the pooling function is $\frac{1}{L+1}$. If we take a closer look at the power of adjacency matrix $\mathbf{\tilde{A}}^l$, we have the following observation:
\begin{equation}
\mathbf{\tilde{A}}^l=
\left\{
             \begin{array}{lr}
             \begin{bmatrix}
 			\left(\mathbf{\tilde{R}}\mathbf{\tilde{R}}^T\right)^{\frac{l}{2}}&\mathbf{0}\\ 
			\mathbf{0} & \left(\mathbf{\tilde{R}}^T\mathbf{\tilde{R}}\right)^{\frac{l}{2}}
			\end{bmatrix}  \quad\quad\quad\quad\,\, l=\{0, 2, 4, \cdots \}    \\\\
             
             \begin{bmatrix}
 			\mathbf{0}& \mathbf{\tilde{R}}\left(\mathbf{\tilde{R}}^T\mathbf{\tilde{R}}
 			\right)^{\frac{l\mbox{-}1}{2}}\\
			\mathbf{R}^T\left(\mathbf{\tilde{R}}\mathbf{\tilde{R}}^T\right)^{\frac{l\mbox{-}1}{2}} &
			\mathbf{0}
			\end{bmatrix}   \quad   l=\{1, 3, 5, \cdots \}.   \\
             \end{array}
\right.
\end{equation}
Following the definition of $\mathbf{\tilde{A}}$, $\mathbf{\tilde{R}}=\mathbf{D}_U^{\mbox{-}\frac{1}{2}}\mathbf{R}\mathbf{D}_I^{\mbox{-}\frac{1}{2}}$, where $\mathbf{D}_U$ and $\mathbf{D}_I$ are the node degree matrices for users and items, respectively. Then, we can split Equation (\ref{lightgcn}) as follows:
\begin{equation}
\begin{aligned}
&\mathbf{O}_U=\frac{\sum_{l=\{0, 2, 4, \cdots\}}\left(\mathbf{\tilde{R}}\mathbf{\tilde{R}}^T \right)^{\frac{l}{2}}\mathbf{E}_U+ \sum_{l=\{1, 3, 5, \cdots\}}\mathbf{\tilde{R}}\left( \mathbf{\tilde{R}}^T\mathbf{\tilde{R}} \right)^{\frac{l\mbox{-}1}{2}}\mathbf{E}_I}{L+1}, \\
&\mathbf{O}_I=\frac{\sum_{l=\{0, 2, 4, \cdots\}}\left( \mathbf{\tilde{R}}^T\mathbf{\tilde{R}} \right)^{\frac{l}{2}}\mathbf{E}_I + \sum_{l=\{1, 3, 5, \cdots\}}\mathbf{\tilde{R}}^T\left( \mathbf{\tilde{R}}\mathbf{\tilde{R}}^T \right)^{\frac{l\mbox{-}1}{2}}\mathbf{E}_U}{L+1}.
\end{aligned}
\label{rewrite_lightgcn}
\end{equation}
The first and second terms represent the messages from homogeneous (even-hops) and heterogeneous (odd-hops) neighborhood, $\mathbf{O}_U$ and $\mathbf{O}_I$ are final embeddings for user and items, $\mathbf{E}_U$ and $\mathbf{E}_I$ are embedding matrices for users and items, respectively. Similar to the definition in Section 2.2, let $\mathbf{P}$, $\mathbf{Q}$, and $\sigma_k$ denote the stacked left, right singular vectors, and singular value for $\mathbf{\tilde{R}}$, respectively, and we formulate the following theorem.

\begin{theorem}
The adjacency relations in Equation (\ref{rewrite_lightgcn}) can be rewritten as the following forms:
\begin{equation}
\begin{aligned}
&\left(\mathbf{\tilde{R}}\mathbf{\tilde{R}}^T\right)^{l}=\mathbf{P}diag\left(\sigma_k^{2l}\right)\mathbf{P}^T,\\
&\left(\mathbf{\tilde{R}}^T\mathbf{\tilde{R}}\right)^{l}=\mathbf{Q}diag\left(\sigma_k^{2l}\right)\mathbf{Q}^T,
\end{aligned}
\label{thorem_11}
\end{equation}
\begin{equation}
\begin{aligned}
&\mathbf{\tilde{R}}\left(\mathbf{\tilde{R}}^T\mathbf{\tilde{R}}\right)^{\frac{l\mbox{-}1}{2}}=\mathbf{P}diag\left(\sigma_k^{l}\right)\mathbf{Q}^T,\\
&\mathbf{R}^T\left(\mathbf{\tilde{R}}\mathbf{\tilde{R}}^T\right)^{\frac{l\mbox{-}1}{2}}=\mathbf{Q}diag\left(\sigma_k^{l}\right)\mathbf{P}^T.
\end{aligned}
\label{thorem_12}
\end{equation} 
\end{theorem}

\begin{figure} \centering 
\subfigure[Odd-hop neighbors.] {  
\includegraphics[width=0.333\columnwidth]{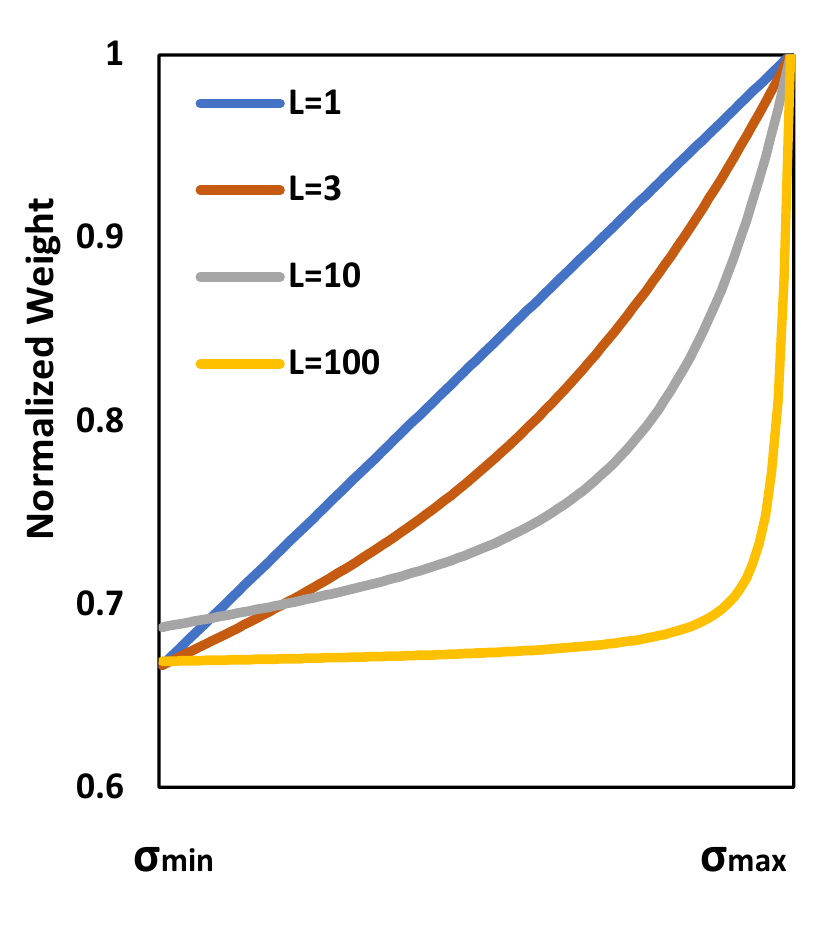} 
}  
\hspace{-0.4cm}
\subfigure[Even-hop neighbors. ] {  
\includegraphics[width=0.333\columnwidth]{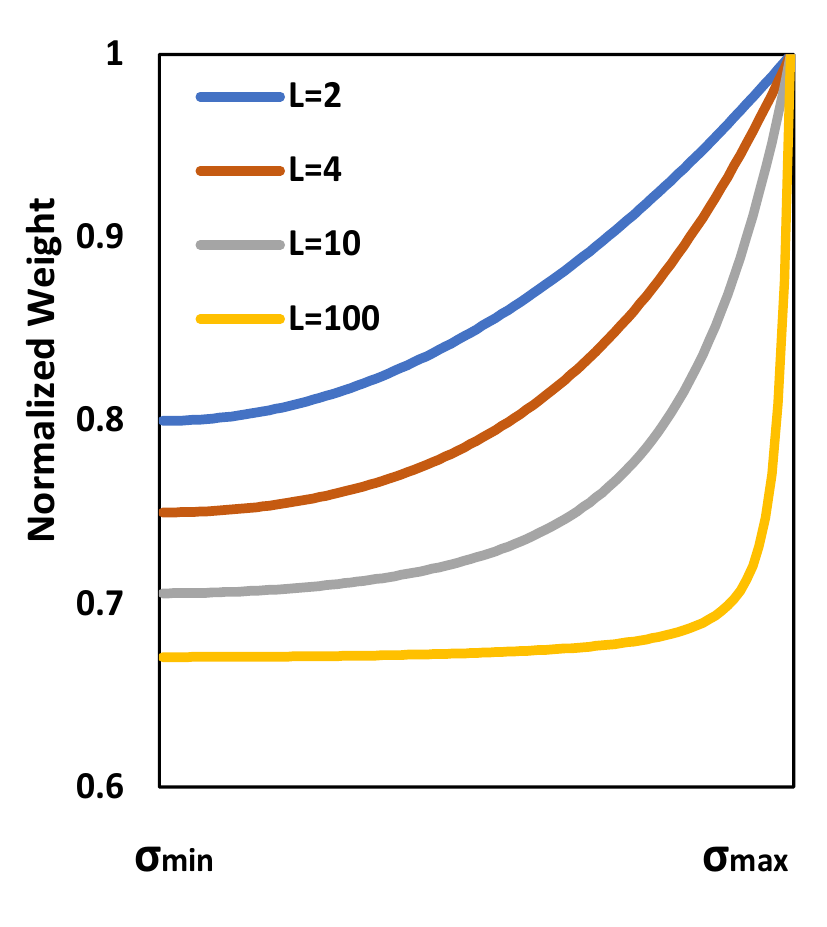} 
}
\hspace{-0.4cm}
\subfigure[SVD-LightGCN.] {  
\includegraphics[width=0.333\columnwidth]{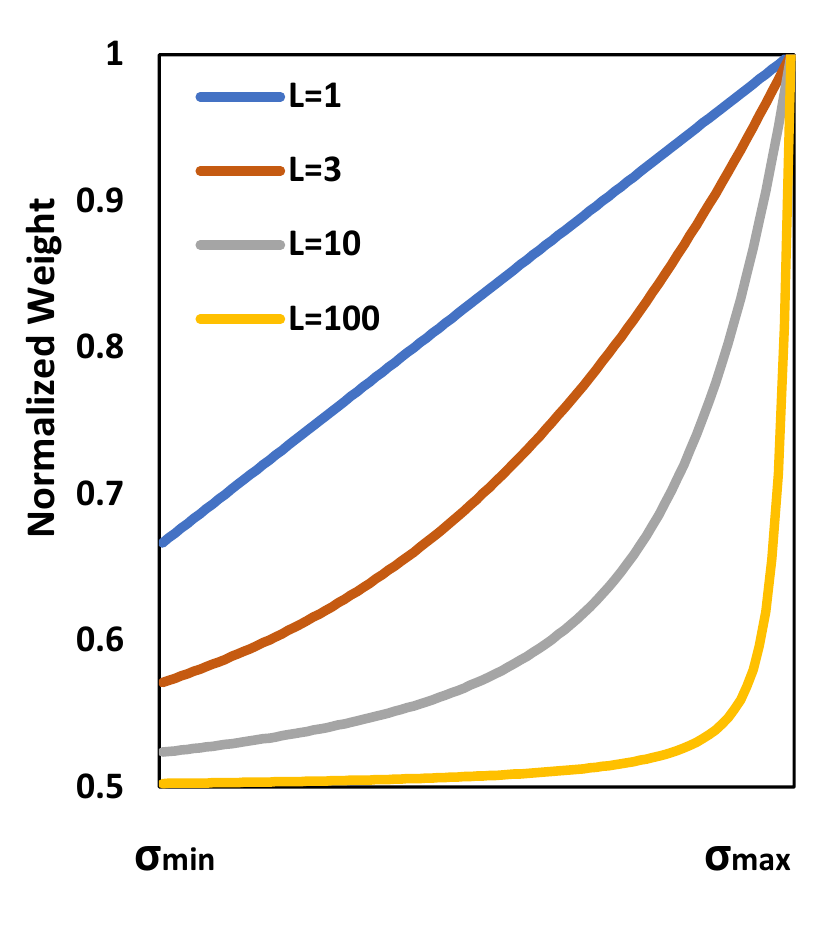} 
}  
\caption{Normalized weights of singular vectors.}  
\label{weight}
\end{figure}

Following Theorem 1, we can rewrite Equation (\ref{rewrite_lightgcn}) as:
\begin{equation}
\begin{aligned}
&\mathbf{O}_U=\mathbf{P}diag\left(\frac{\sum_{l=\{0, 2, \cdots \}}\sigma_k^l}{L+1}\right)\mathbf{P}^T\mathbf{E}_U + \mathbf{P}diag\left(\frac{\sum_{l=\{1, 3, \cdots \}}\sigma_k^l}{L+1}\right)\mathbf{Q}^T\mathbf{E}_I,\\
&\mathbf{O}_I=\mathbf{Q}diag\left(\frac{\sum_{l=\{0, 2, \cdots \}}\sigma_k^l}{L+1}\right)\mathbf{Q}^T\mathbf{E}_I + \mathbf{Q}diag\left(\frac{\sum_{l=\{1, 3, \cdots \}}\sigma_k^l}{L+1}\right)\mathbf{P}^T\mathbf{E}_U.
\end{aligned}
\label{simplify_1}
\end{equation}
Now the final embeddings are contributed from $\mathbf{\tilde{R}}$'s singular vectors and values instead of neighborhood. Note that:
\begin{equation}
\mathbf{P}diag\left(\frac{\sum_{l=\{0, 2, \cdots \}}\sigma_k^l}{L+1}\right)\mathbf{P}^T=\sum_k \frac{\sum_{l=\{0, 2, \cdots \}}\sigma_k^l}{L+1} \mathbf{p}_k \mathbf{p}^T_k.
\end{equation} 
$\frac{\sum_{l=\{0, 2, \cdots \}}\sigma_k^l}{L+1}$ and $\frac{\sum_{l=\{1, 3, \cdots \}}\sigma_k^l}{L+1}$ can be considered as weights of singular vectors when considering even and odd hop neighbors, respectively. We illustrate the normalized weights in Figure \ref{weight} (a) and (b), and make the following observation:

\begin{observation}
As stacking more graph convolution layers, the goal of GCNs is to learn a low-rank representation by stressing (suppressing) more components with larger (smaller) singular values.
\end{observation}

We further observe that:
\begin{equation}
\mathbf{O}_u=\left(\mathbf{p}_{u*}^T\odot\frac{\sum_{l=\{0, 2, \cdots \}}\sigma^l}{L+1} \right)\mathbf{P}^T \mathbf{E}_U + \left(\mathbf{p}_{u*}^T\odot\frac{\sum_{l=\{1, 3, \cdots \}}\sigma^l}{L+1} \right)\mathbf{Q}^T \mathbf{E}_I, 
\end{equation}
where $\sigma$ is a vector containing all singular values, $\mathbf{p}_{u*}^T$ is the $u$-th row vector, $\odot$ represents the element-wise multiplication. We can see $\mathbf{P}^T\mathbf{E}_U$ and $\mathbf{Q}^T\mathbf{E}_I$ are common terms for distinct users/items, what makes representations unique lies in the term in parentheses. 
\begin{assumption}
$\mathbf{P}^T\mathbf{E}_U$ and $\mathbf{Q}^T\mathbf{E}_I$ are redundant.
\end{assumption}
On the other hand, the above two terms play a important role constituting the core design of GCNs (i.e., neighborhood aggregation), replacing or removing them leads to a new learning paradigm without explicitly aggregating neighborhood. To verify this assumption, we evaluate three models: (1) the original model Equation (\ref{simplify_1}); (2) we simply replace $\mathbf{P}^T\mathbf{E}_U$ and $\mathbf{P}^T\mathbf{E}_I$ with two different weight matrices; (3) we use a shared weight matrix based on (2).

The results in Figure \ref{analysis} (a) and (b) show that the performance of the three models are fairly close, and thus: (1) neighborhood aggregation is not necessary for GCNs; (2) The power of GCNs for CF does not heavily rely on model parameters, since reducing parameters (by half) does not reduce the accuracy and even results in faster convergence. Based on the model (3), we can merge the two terms in Equation (\ref{simplify_1}) and simplify it as:  
\begin{equation}
\begin{aligned}
&\mathbf{O}_U=\mathbf{P}diag\left(\frac{\sum_{l=0}^L \sigma_k^l}{L+1}\right)\mathbf{W},\\
&\mathbf{O}_I=\mathbf{Q}diag\left(\frac{\sum_{l=0}^L \sigma_k^l}{L+1}\right)\mathbf{W},
\end{aligned}
\label{simplify_2}
\end{equation} 
and name it SVD-LightGCN. We can interpret it as a two-step procedure. We first obtain a weighted singular matrices by assigning the weight $\frac{\sum_{l=0}^L \sigma_k^l}{L+1}$ to singular vectors (i.e., $\mathbf{p}_k$ and $\mathbf{q}_k$); then, we learn a condensed embeddings of the singular vectors through a feature transformation $\mathbf{W}$. Figure \ref{weight} (c) shows the goal of SVD-LightGCN is also to learn a low-rank representation, where the weights of singular vectors are adjustable through $L$. We also observe that:
\begin{observation}
SVD is a special case of SVD-LightGCN where $\mathbf{W}=\mathbf{I}$ and $l=L=\frac{1}{2}$ (fixed to a square root).
\end{observation}

\subsection{Analysis on SVD-LightGCN}
\textbf{Training Efficiency.} Observation 1 provides an alternative way to build GCNs, that we can directly focus on the weights over singular vectors instead of stacking layers. However, retrieving all singular vectors is computationally expensive and not applicable on large datasets as well. On the other hand, Observation 1 implies that most small singular values are not so helpful for recommendation. To further verify this observation, we compare SVD-LightGCN and SVD-LightGCN-T which only exploits $K$-largest singular values and vectors, and report the accuracy of them in Figure \ref{analysis} (c) and (d), where x-axis represents the singular vector ratio: $\frac{K}{{\rm min}(\left|\mathcal{U}\right|,\left|\mathcal{I}\right|)}$. We can see SVD-LightGCN-T with only the top 1\% largest singular values and vectors outperforms SVD-LightGCN which exploits all singular vectors, and the best accuracy is achieved at 4\% on CiteULike, 6\% on ML-100K. This finding not only shows that most small singular values and vectors are noisy that even reduces the performance, but also  helps largely reduce the training cost and improve the training efficiency. For instance, retrieving 4\% of the singular vectors and values only takes 1.8s on CiteULike, the learning parameters ($Kd$) are merely 1\% of that of MF and LightGCN ($\left|\mathcal{U}\right|d+\left|\mathcal{I}\right|d$).\\\\

\textbf{Over-Smoothing.} Users and items tend to have the same representations when the model layer $L$ is large enough \cite{li2018deeper}.
\begin{theorem}
The maximum singular value of $\mathbf{\tilde{R}}$ is 1.
\end{theorem}

As shown from Figure \ref{weight} (b), the larger singular values are further emphasized as increasing the model layers. Following Theorem 2, if we further increase the layer $L$:
\begin{equation}
\lim_{ L\rightarrow\infty}=\frac{\sum_{l=0}^{L}\frac{\sigma_k^l}{L+1}}{\sum_{l=0}^{L}\frac{\sigma_{{\rm max}}^l}{L+1}}\rightarrow0,
\end{equation}       
where the weights of any singular vectors are reduced to 0 compared with the largest one $\sigma_{{\rm max}}$, where user/item representations are only contributed by the largest singular vector. Thus, increasing model layers does not necessarily lead to better representations and might instead cause information loss. The over-smoothing issue lies in the gap between singular values, where it is enlarged as stacking layers, which suppresses some important information that matters for recommendation. To alleviate this issue, we define a renormalized interaction matrix as: $\dot{\mathbf{R}}=(\mathbf{D}_U+\alpha\mathbf{I})^{\mbox{-}\frac{1}{2}}\mathbf{R}(\mathbf{D}_I+\alpha\mathbf{I})^{\mbox{-}\frac{1}{2}}$ where $\alpha\geq0$.
\begin{theorem}
Given the singular value $\dot{\sigma}_k$ of $\dot{\mathbf{R}}$, $\dot{\sigma}_{\rm{max}} \leq \frac{d_{{\rm max}}}{d_{{\rm max}+\alpha}}$ where $d_{{\rm max}}$ is the maximum node degree.
\end{theorem}
  
The maximum singular value becomes smaller as increasing $\alpha$, indicating a smaller gap. On the other hand, a too small gap fails to emphasize the difference of importance of different components (i.e., the component with a larger singular value is more important). Thus, we can adjust $\alpha$ to regulate the gap to assure that important information is well preserved and to adapt to different datasets. \par
Furthermore, the weighting function is a crucial design as it controls the weights of singular vectors, while LightGCN adopts a polynomial in a heuristic way. Let $\psi(\cdot)$ denotes the weighting function. Basically, we can parameterize $\psi(\cdot)$ with advanced algorithms to dynamically learn the weights of singular vectors. Alternatively, if we consider $\psi(\cdot)$ as a static continuous function of singular values $\sigma_k$, it is expected to weight the singular vectors through a function with easy-to-adjust hyperparameters instead of by repeatedly increasing the model layer $L$. In addition, by replacing the polynomial in LightGCN with $\psi(\cdot)$, following the Taylor series $\psi(\sigma_k)=\sum^L_{l=0}\alpha_l \sigma^l_k$, we can rewrite Equation (\ref{lightgcn}) as:
\begin{equation}
\mathbf{O}=\left(\sum_{l=0}^L\alpha_l \mathbf{\tilde{A}}^l \right) \mathbf{E},
\label{lightgcn_back}
\end{equation}       
where $\alpha_l$ is $\psi(\sigma_k)$'s $l$-th order derivative at 0, $L$ is $\psi(\cdot)$'s highest order. From a spatial perspective, $\alpha_l$ is also the contribution of $l$-th order neighborhood, and $L$ corresponds to the farthest neighborhood being incorporated. Intuitively, it is expected that user/item representations are constructed from as many positive neighborhood signals as possible (i.e., $\alpha_l>0$ and $L\rightarrow\infty$), implying that $\psi(\cdot)$ is infinitely differentiable with any-order derivatives positive.

\subsection{SVD-GCN} 
Based on the analysis in Section 3.2, we formulate the user and item representations as follows:
\begin{equation}
\begin{aligned}
&\mathbf{O}_U=\dot{\mathbf{P}}^{(K)}diag\left(\psi\left(\dot{\sigma_k}\right)\right)\mathbf{W},\\
&\mathbf{O}_I=\dot{\mathbf{Q}}^{(K)}diag\left(\psi\left(\dot{\sigma_k}\right)\right)\mathbf{W},
\end{aligned}
\label{svd_gcn_matrix}
\end{equation}
where $\dot{\mathbf{P}}^{(K)}$ and $\dot{\mathbf{Q}}^{(K)}$ are composed of $K$-largest left and right singular vectors of $\dot{\mathbf{R}}$, respectively. Our initial attempt is to dynamically model the importance of singular vectors through a neural network given singular values as the input. However, we found that such a design underperforms static designs in most cases, and speculate that the reason is due to the data sparsity on CF. Unlike other recommendation tasks with rich side information, the only available data is the user/item ID besides interactions, which increases the difficulty to learn the intrinsic data characteristics. Based on previous analysis in Section 3.2, extensive experiments show that an exponential kernel \cite{Kondor2002diffusion} achieves superior accuracy on the tested data, thus we set $\psi(\dot{\sigma_k})=e^{\beta\dot{\sigma_k}}$, where $\beta$ is a hyperparameter to adjust the extent of emphasis over larger singular values (i.e., a larger (smaller) $\beta$ emphasizes the importance of larger (smaller) singular values more). We will also compare different $\psi(\cdot)$ designs in Section 4.3.  Unlike conventional GCNs updating all user/item embeddings simultaneously in a matrix form resulting in a large spatial complexity, we can train SVD-GCN in a node form with more flexibility as:
\begin{equation}
\begin{aligned}
&\mathbf{o}_u=\dot{\mathbf{p}}^T_u \odot \left(e^{\beta\dot{\sigma}}\right)\mathbf{W},\\
&\mathbf{o}_i=\dot{\mathbf{q}}^T_i \odot \left(e^{\beta\dot{\sigma}}\right)\mathbf{W},
\end{aligned}
\label{svd_gcn_node}
\end{equation} 
where $\dot{\mathbf{p}}^T_u$ and $\dot{\mathbf{q}}^T_i$ are the rows of $\dot{\mathbf{P}}^{(K)}$ and $\dot{\mathbf{Q}}^{(K)}$, respectively; $\dot{\sigma}$ is a vector containing all singular values. Note that the element-wise multiplication does not involve parameters thus can be precomputed. Then, inspired by BPR loss \cite{rendle2009bpr}, we formulate the loss function as follows:
\begin{equation}
\mathcal{L}_{main}=\sum_{u\in\mathcal{U}}\sum_{(u,i^+)\in\mathbf{R}^+,(u,i^{\mbox{-}})\notin\mathbf{R}^+}\ln \sigma \left(\mathbf{o}_u^T\mathbf{o}_{i^+} - \mathbf{o}_u^T\mathbf{o}_{i^{\mbox{-}}} \right).
\label{svd_gcn_base}
\end{equation}
As shown in Equation (\ref{rewrite_lightgcn}), in GCN-based CF methods, user/item representations are contributed from three kinds of information flows: user-item, user-user, and item-item relations. Thus, besides the user-item relations, homogeneous (i.e., user-user and item-item) relations also help increase model effectiveness. We define a user-user $\mathcal{G}_U=(\mathcal{V}_U, \mathcal{E}_U)$, and an item-item graph $\mathcal{G}_I=(\mathcal{V}_I, \mathcal{E}_I)$, where $\mathcal{V}_U=\mathcal{U}$ and $\mathcal{V}_I=\mathcal{I}$; $\mathcal{E}_U=\{(u, g)|g\in\mathcal{N}_i, i\in\mathcal{N}_u\}$ and $\mathcal{E}_I=\{(i, h)|h\in\mathcal{N}_u, u\in\mathcal{N}_i\}$, where $\mathcal{N}_u$ and $\mathcal{N}_i$ are the sets of directly connected neighbors for $u$ and $i$, respectively. Naturally, we can define the normalized adjacency matrix of $\mathcal{G}_U$ and $\mathcal{G}_I$ as $\mathbf{R}_U=\dot{\mathbf{R}}^T\dot{\mathbf{R}}$ and $\mathbf{R}_I=\dot{\mathbf{R}}\dot{\mathbf{R}}^T$, respectively. According to Equation (\ref{proof_1}) in Section 7, the eigenvectors of $\mathbf{R}_U$ and $\mathbf{R}_I$ are actually $\dot{\mathbf{R}}$'s left and right singular vectors, respectively; and the eigenvalues are both the square of $\dot{\mathbf{R}}$'s singular values. Thus, $\mathcal{G}$, $\mathcal{G}_U$ and $\mathcal{G}_I$ are closely connected. We formulate the following loss to learn the relations on $\mathcal{G}_U$:
\begin{equation}
\mathcal{L}_{user}=\sum_{u\in\mathcal{U}}\sum_{(u, u^+)\in\mathcal{E}_U, (u, u^{\mbox{-}})\notin\mathcal{E}_U}\ln \sigma \left(\mathbf{o}_u^T\mathbf{o}_{u^+} - \mathbf{o}_u^T\mathbf{o}_{u^{\mbox{-}}} \right).
\label{svd_gcn_user}
\end{equation}          
Similarly, we learn the relations on $\mathcal{G}_I$ via the following loss:
\begin{equation}
\mathcal{L}_{item}=\sum_{i\in\mathcal{I}}\sum_{(i, i^+)\in\mathcal{E}_I, (i, i^{\mbox{-}})\notin\mathcal{E}_I}\ln \sigma \left(\mathbf{o}_i^T\mathbf{o}_{i^+} - \mathbf{o}_i^T\mathbf{o}_{i^{\mbox{-}}} \right).
\label{svd_gcn_item}
\end{equation}
Finally, we propose the following four SVD-GCN variants:
\begin{equation}
\begin{aligned}
&\rm{SVD\mbox{-}GCN\mbox{-}B}: \; \mathcal{L}=\mathcal{L}_{main}+\lambda\left \| \Theta \right \|^2_2,\\
&\rm{SVD\mbox{-}GCN\mbox{-}U}: \; \mathcal{L}=\mathcal{L}_{main}+\gamma\mathcal{L}_{user}+\lambda\left \| \Theta \right \|^2_2,\\
&\rm{SVD\mbox{-}GCN\mbox{-}I}: \; \mathcal{L}=\mathcal{L}_{main}+\zeta\mathcal{L}_{item}+\lambda\left \| \Theta \right \|^2_2,\\
&\rm{SVD\mbox{-}GCN\mbox{-}M}: \; \mathcal{L}=\mathcal{L}_{main}+\gamma\mathcal{L}_{user}+\zeta\mathcal{L}_{item}+\lambda\left \| \Theta \right \|^2_2,\\
\end{aligned}
\label{final_svd_gcn}
\end{equation}
where $\Theta$ denotes the model parameters. Besides the above variants, to evaluate the effect of the feature transformation, we propose a non-parametric method SVD-GCN-S by removing $\mathbf{W}$.
\subsection{Discussion}
\subsubsection{Model Complexity}
The complexity of SVD-GCN mainly comes from two parts. We first retrieve $K$ singular vectors through SVD for the low-rank matrix \cite{halko2011finding}, with a complexity as: $\mathcal{O}(K\left|\mathbf{R}^+\right|+K^2\left|\mathcal{U}\right|+K^2\left|\mathcal{I} \right|)$. We run the algorithm on GPU and only require a very few singular vectors, which only costs several seconds. Except for SVD-GCN-S, other variants require training with time complexity as $\mathcal{O}(c\left|\mathbf{R}^+\right|(K+1)d)$, which is comparable to MF: $c\left|\mathbf{R}^+\right|d$, where $c$ denotes the number of epochs. On the other hand, the model parameters of MF is $\frac{\left|\mathcal{U}\right|+\left|\mathcal{I}\right|}{K}$ time that of GCN-SVD. Overall, SVD-GCN is lighter than MF, and we will show more quantitative results in terms of efficiency in Section 4.2.

\subsubsection{Comparison with GCN-based CF Methods}  
Compared with conventional GCN-based methods, GCN-SVD replaces neighborhood aggregation with a truncated SVD and significantly reduces the model parameters. Overall, SVD-GCN is equipped with a lighter structure and more scalable. Recent proposed work UltraGCN \cite{mao2021ultragcn} simplifies LightGCN by replacing neighborhood aggregation with a weighted MF and shows lower complexity:
\begin{equation}
{\rm max}\sum_{u\in\mathcal{U}, i\in\mathcal{N}_u} \beta_{u,i}\mathbf{e}_u^T\mathbf{e}_i,
\label{ultragcn}
\end{equation}
where $\beta_{u,i}$ is obtained from single-layer LightGCN. However, UltraGCN improves based on single-layer LightGCN, which can only exploit the first order neighborhood and losses the ability of incorporating high-order neighborhood to augment training interactions. On the other hand, SVD-GCN is derived from any-layer LightGCN and we further generalize it to the situation of infinite layers, hence maximizes the power of GCNs.  

\begin{table}
\centering
\caption{Statistics of datasets}
\begin{tabular}{lcccc}
\toprule
Datasets&\#User&\#Item &\#Interactions &Density\%\\
\midrule
CiteULike&5,551&16,981&210,537&0.223\\
ML-100K&943&1,682&100,000&6.305\\
\midrule
ML-1M&6,040&3,952&1,000,209&4.190\\
Yelp&25,677&25,815&731,672&0.109\\
Gowalla&29,858&40,981&1,027,370&0.084\\
\bottomrule
\label{datasets}
\end{tabular}
\end{table}

\section{Experiments}
In this section, we comprehensively evaluate our proposed SVD-GCN. The rest of this section is organized as follows: we introduce experimental settings in Section 4.1, compare baselines with SVD-GCN in terms of recommendation accuracy and training efficiency in Section 4.2; in Section 4.3, we dissect SVD-GCN to show the effectiveness of our proposed designs and how different hyperparameter settings (i.e., $K$, $\alpha$, $\beta$, $\gamma$, and $\zeta$) affect performance.
\subsection{Experimental Settings}
\subsubsection{Datasets and Evaluation Metrics}
We use five public datasets in this work, where the results of Figure \ref{analysis} are based on CiteULike\footnote{\url{https://github.com/js05212/citeulike-a}} and ML-100K \cite{harper2015movielens}. To demonstrate the effectiveness of our proposed methods on more datasets and to justify the previous analysis, we evaluate SVD-GCN on three other datasets: Gowalla \cite{wang2019neural}, Yelp \cite{he2016fast}, and ML-1M \cite{harper2015movielens}. Since we focus on implicit feedback, we only keep user/item ID and transform feedbacks to binary ratings. Table \ref{datasets} lists statistics of datasets.\par 
We adopt two widely-used metrics: Recall and nDCG \cite{jarvelin2002cumulated} to evaluate our methods. Recall measures the ratio of the relevant items in the recommended list to all relevant items in test sets, while nDCG takes the ranking into consideration by assigning higher scores to  items ranking higher. The recommendation list is generated by ranking unobserved items and truncating at position $k$. Since the advantage of GCN-based methods over traditional CF methods is the ability of leveraging high-order neighborhood to augment training data, thereby alleviating the data sparsity, we only use 20\% of interactions for training and leave the remaining for test to evaluate the model robustness and stability; we randomly select 5\% from the training set as validation set for hyper-parameter tuning and report the average accuracy on test sets.

\subsubsection{Baselines}
We compare our methods with the following competing baselines, where the hyperparameter settings are based on the results of the original papers:
\begin{itemize}[leftmargin=10pt]

\item BPR \cite{rendle2009bpr}: This is a stable and classic MF-based method, exploiting a Bayesian personalized ranking loss for personalized rankings.

\item EASE \cite{steck2019embarrassingly}: This is a neighborhood-based method with a closed form solution and show superior performance to many traditional CF methods. 

\item LightGCN \cite{he2020lightgcn}: This method uses a light GCN architecture for CF by removing activations functions and feature transformation. We use a three-layer architecture as the baseline.

\item LCFN \cite{yu2020graph}: This model replaces the original graph convolution with a low pass graph convolution to remove the noise from interactions for recommendation. We set $F=0.005$ and use a single-layer architecture.

\item SGL-ED \cite{wu2020self}: This model generates different node views by randomly removing the edge connections and maximizes their agreements, and the proposed self-supervised loss is implemented on LightGCN \cite{he2020lightgcn}. We set $\tau=0.2$, $\lambda_1=0.1$, $p=0.1$, and use a three-layer architecture.

\item UltraGCN \cite{mao2021ultragcn}: This model simplifies LightGCN by replacing neighborhood aggregation with a weighted MF, which shows faster convergence and less complexity. 

\end{itemize} 
We remove some popular GCN-based methods such as Pinsage \cite{ying2018graph}, NGCF \cite{wang2019neural}, and SpectralCF \cite{zheng2018spectral} as aforementioned baselines have already shown superiority over them. 

\subsubsection{Implementation Details} 
We implemented the proposed model based on PyTorch$\footnote{\url{https://pytorch.org/}}$ and released the code on Github$\footnote{\url{https://github.com/tanatosuu/svd_gcn}}$. For all models, We use SGD as the optimizer, the embedding size $d$ is set to 64, the regularization rate $\lambda$ is set to 0.01 on all datasets, the learning rate is tuned amongst $\{0.001,0.005,0.01,\cdots,1\}$; without specification, the model parameters are initialized with Xavier Initialization \cite{glorot2010understanding}; the batch size is set to 256. We report other hyperparameter settings in the next subsection.

\begin{table*}[]
\caption{Overall performance comparison.}
\begin{tabular}{c|cccc|cccc|cccc}
\hline
              & \multicolumn{4}{c|}{\textbf{Yelp}}                                               & \multicolumn{4}{c|}{\textbf{ML-1M}}                                              & \multicolumn{4}{c}{\textbf{Gowalla}}                                            \\
              & \multicolumn{2}{c}{\textbf{nDCG@$k$}} & \multicolumn{2}{c|}{\textbf{Recall@$k$}} & \multicolumn{2}{c}{\textbf{nDCG@$k$}} & \multicolumn{2}{c|}{\textbf{Recall@$k$}} & \multicolumn{2}{c}{\textbf{nDCG@$k$}} & \multicolumn{2}{c}{\textbf{Recall@$k$}} \\
              & \textbf{$k$=10}   & \textbf{$k$=20}   & \textbf{$k$=10}     & \textbf{$k$=20}    & \textbf{$k$=10}   & \textbf{$k$=20}   & \textbf{$k$=10}     & \textbf{$k$=20}    & \textbf{$k$=10}   & \textbf{$k$=20}   & \textbf{$k$=10}    & \textbf{$k$=20}    \\ \hline
BPR           & 0.0388            & 0.0374            & 0.0371              & 0.0370             & 0.5521            & 0.4849            & 0.5491              & 0.4578             & 0.1086            & 0.0907            & 0.0917             & 0.0743             \\
Ease          & 0.0360            & 0.0362            & 0.0346              & 0.0368             & 0.3773            & 0.3249            & 0.3682              & 0.3000             & 0.0722            & 0.0670            & 0.0680             & 0.0642             \\ \hline
LCFN          & 0.0617            & 0.0627            & 0.0613              & 0.0653             & 0.5927            & 0.5197            & 0.5887              & 0.4898             & 0.1305            & 0.1132            & 0.1144             & 0.0980             \\
UltraGCN      & 0.0417            & 0.0403            & 0.0404              & 0.0403             & 0.5326            & 0.4688            & 0.5302              & 0.4434             & 0.0977            & 0.0815            & 0.0841             & 0.0681             \\
LightGCN      & 0.0679            & 0.0680            & 0.0669              & 0.0704             & 0.5917            & 0.5261            & 0.5941              & 0.5031             & 0.1477            & 0.1327            & 0.1368             & 0.1224             \\
SGL-ED        & \underline{0.0817}      & \underline{0.0794}      & \underline{0.0784}        & \underline{0.0792}       & \underline{0.6029}      & \underline{0.5314}      & \underline{0.6010}        & \underline{0.5035}       & \underline{0.1789}      & \underline{0.1561}      & \underline{0.1563}       & \underline{0.1353}       \\ \hline
SVD-GCN-S     & 0.0919            & 0.0895            & 0.0894              & 0.0903 & 0.6458            & 0.5702            & 0.6466              & 0.5421             & 0.1900            & 0.1677            & 0.1690             & 0.1484             \\
SVD-GCN-B     & 0.0898            & 0.0876            & 0.0866              & 0.0879             & 0.6480            & 0.5724            & 0.6484              & 0.5443             & 0.1820            & 0.1607            & 0.1628             & 0.1428             \\
SVD-GCN-U     & 0.0923            & 0.0897            & 0.0888              & 0.0898             & 0.6571   & \textbf{0.5791}   & \textbf{0.6571}     & \textbf{0.5495}    & 0.1875 & 0.1654            & 0.1667             & 0.1460             \\
SVD-GCN-I     & 0.0930   & 0.0907   & 0.0897     & 0.0910    & \textbf{0.6574}            & 0.5770            & 0.6565              & 0.5465             & 0.1857            & 0.1646            & 0.1662             & 0.1466             \\
SVD-GCN-M     & \textbf{0.0941}            &\textbf{ 0.0917 }           & \textbf{0.0908             } & \textbf{0.0921} & 0.6521            & 0.5705            & 0.6490              & 0.5377             & \textbf{0.1905}   & \textbf{0.1681}   & \textbf{0.1693}    & \textbf{0.1487}    \\ \hline
Improv.\% & +15.18            & +15.49            & +15.82              & +16.29             & +9.04             & +8.98             & +9.33               & +9.14              & +6.48 & +7.69             & +8.32              & +9.90              \\ \hline
\end{tabular}
\label{per_comparison}
\end{table*}

\begin{table}[]
\caption{Training time comparison on Gowalla.}
\scalebox{0.89}{
\begin{tabular}{c|c|c|c|c}
\hline
Model     & Time/Epoch & Epochs  & Running Time  &Parameters \\ \hline
LightGCN  & 6.43s       & 600        & 3,858s     & 4.5m  \\
UltraGCN  & 2.55s       & 90         & 229.5s    & 4.5m   \\
BPR       & 1.04s       & 250        & 260.0s    & 4.5m   \\ \hline
SVD-GCN-S & 0.00s       & 0         & 3.07s    & 0.0k  \\
SVD-GCN-B & 1.28s       & 8         & 13.31s    & 5.7k  \\
SVD-GCN-U & 2.06s       & 8         & 19.55s    & 5.7k   \\
SVD-GCN-I & 2.18s       & 8         & 20.51s    & 5.7k   \\
SVD-GCN-M & 3.05s      & 8         & 27.47s     & 5.7k  \\ \hline
\end{tabular}}
\label{effi_comparison}
\end{table}

\subsection{Comparison}

\subsubsection{Performance}
We report the accuracy of baselines and our proposed GCN-SVD variants in Table \ref{per_comparison}, and have the following observations:
\begin{itemize}[leftmargin=10pt]

\item Overall, GCN-based methods outperforms traditional CF methods, indicating the effectiveness of GCNs for CF and demonstrating the importance of augmenting training interactions by incorporating high-order neighborhood information, thereby alleviating data sparsity.

\item Among all baselines, SGL-ED achieves the best across all datasets, while our proposed SVD-GCNs show consistent improvements over SGL-ED, indicating the effectiveness and superiority over conventional GCN designs. UltraGCN shows relatively poor performance among GCN-based methods. As shown in our previous analysis in Section 3.4.2, UltraGCN improves based on single-layer GCN which fails to leverage the higher-order neighborhood, thus cannot perform stably with limited interactions.    

\item Since our key contribution is to replace neighborhood aggregation, the improvement is more clear if we compare with pure GCN-based methods such as LightGCN. SVD-GCN outperforms LightGCN on Yelp, ML-1M, and Gowalla by 53.6\%, 11.7\%, and 29.0\%, respectively, in terms of nDCG@10. The improvements over sparse data tend to be more significant, indicating the stability of SVD-GCN under extreme data sparsity.

\item Among SVD-GCN variants, the basic model SVD-GCN-B and SVD-GCN-S already outperform all baselines by a large margin. In addition, introducing user-user and item-item relations results in further improvement. We also notice that mixing user-user and item-item relations does not necessarily leads to better accuracy, and we speculate that the reason might be related to the data density. On the dense data such as ML-1M where the user-item interactions are relatively sufficient, the improvement by introducing user-user and item-item relations is not as significant as that of sparser datasets, and incorporating both relations even performs worse; while on the sparest data Gowalla, introducing auxiliary relations shows consistent improvements.  

\end{itemize}

\subsubsection{Training Efficiency}
\begin{figure} \centering 
\subfigure[Gowalla] {  
\includegraphics[width=0.333\columnwidth]{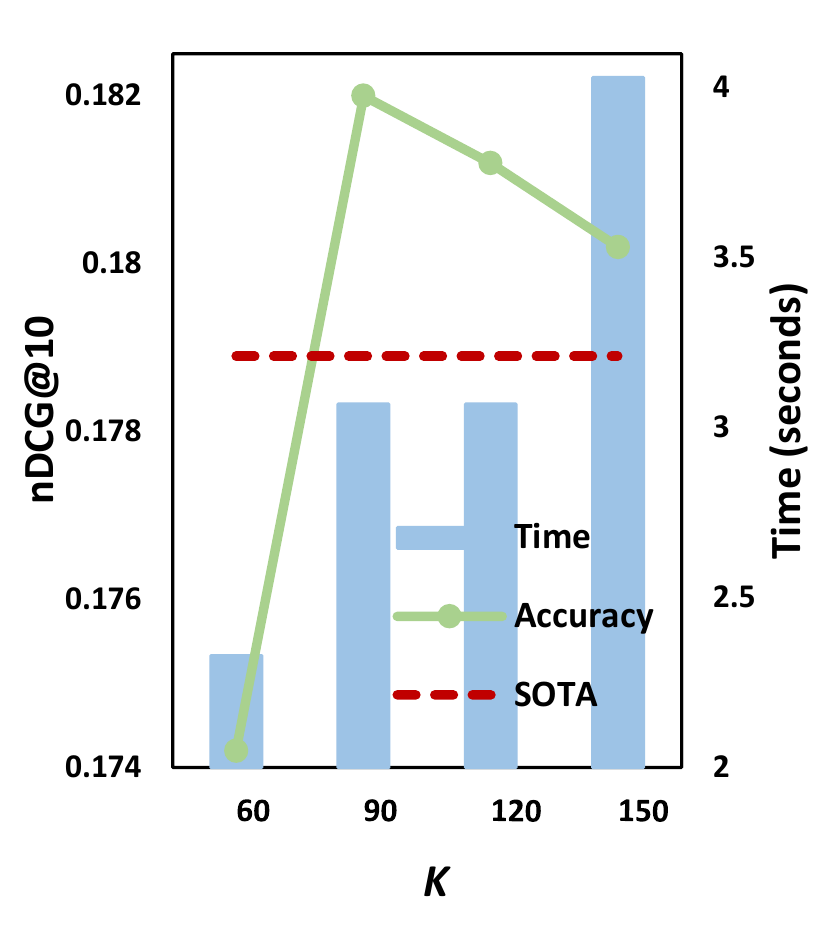} 
}
\hspace{-0.4cm}
\subfigure[Yelp. ] {  
\includegraphics[width=0.333\columnwidth]{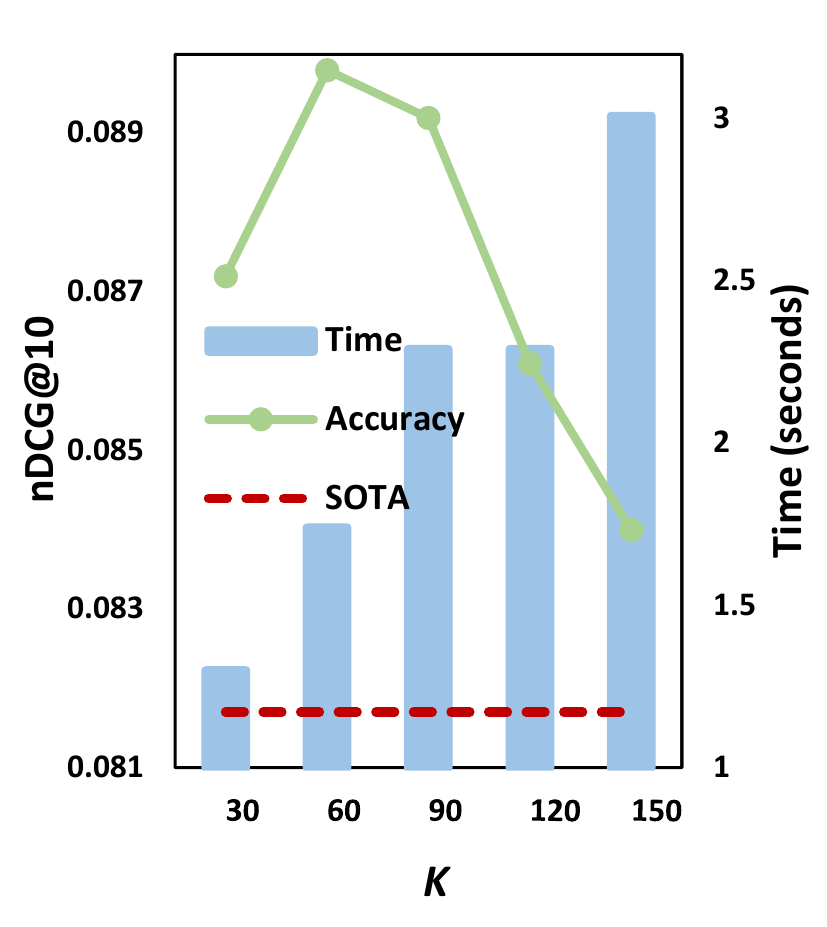} 
}
\hspace{-0.4cm}
\subfigure[MovieLens. ] {  
\includegraphics[width=0.333\columnwidth]{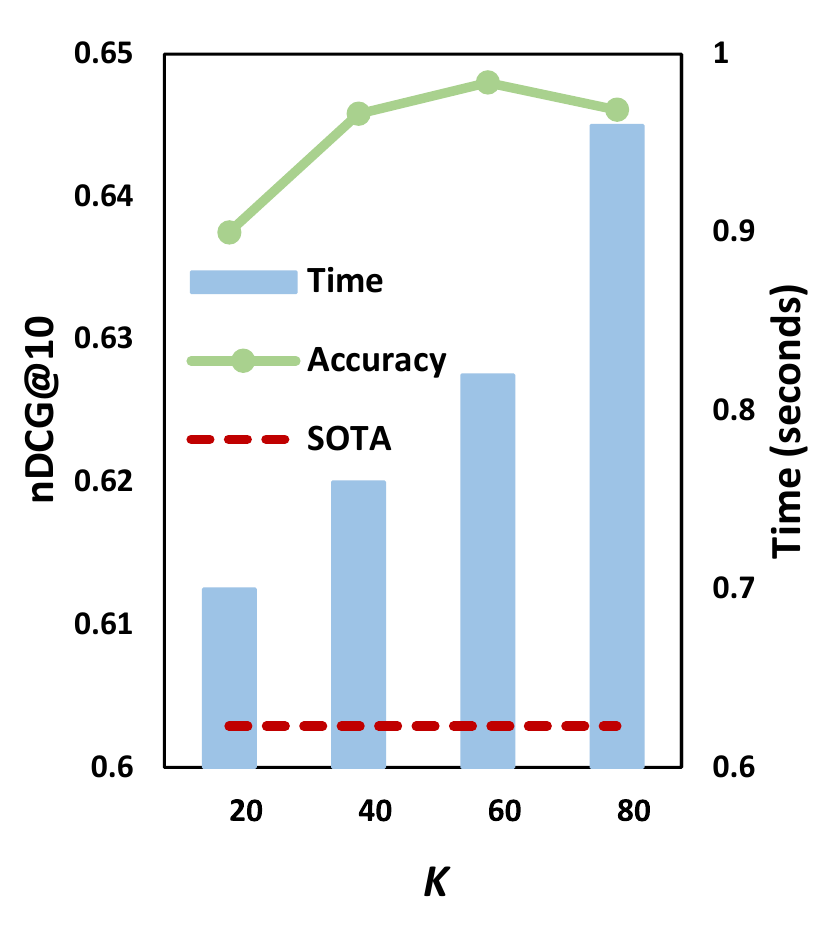} 
}
\vspace{-0.4cm}
\caption{How the preprocessing time and accuracy (nDCG@10) vary on $K$ on SVD-GCN-B.}
\label{preprocess}
\end{figure}

The results shown in this subsection are obtained on a machine equipped with AMD Ryzen 9 5950X and GeForce RTX 3090. Figure \ref{preprocess} shows how the preprocessing time and accuracy change with $K$, where SOTA is the best baseline. The best accuracy is achieved at $K=90$, $K=60$, and $K=60$, where the preprocessing time is 3.07s, 0.82s, and 1.74s, on Gowalla, ML-1M, and Yelp, respectively. Overall, only 1\% singular vectors are required on ML-1M, and less than 0.5\% singular vectors are required on Gowalla and Yelp, when the model reaches the best accuracy.\par

Table \ref{effi_comparison} shows the training time and running epochs of several methods, where the running time includes both preprocessing and training time. Overall, LightGCN is the most time consuming model (3,858s) as it is a conventional GCN model; SVD-GCN-S is the most time efficient model (3.07s) since it does not require model optimization and shows over 1000x speed-up over LightGCN. BPR is the fastest model (1.04s) in terms of training time per epoch, while it still requires hundreds epochs to reach the best accuracy due to the large amount of parameters need to be optimized. Although SVD-GCN variants (excluding SVD-GCN-S) are slightly slower than BPR on training time per epoch, they show fast training convergence as the model parameters are only 0.08\% of that of BPR.

\subsection{Model Analysis}

\subsubsection{How Homogeneous Relations Affect Performance?}
\begin{figure} \centering 
\subfigure[Yelp.] {  
\includegraphics[width=0.45\columnwidth]{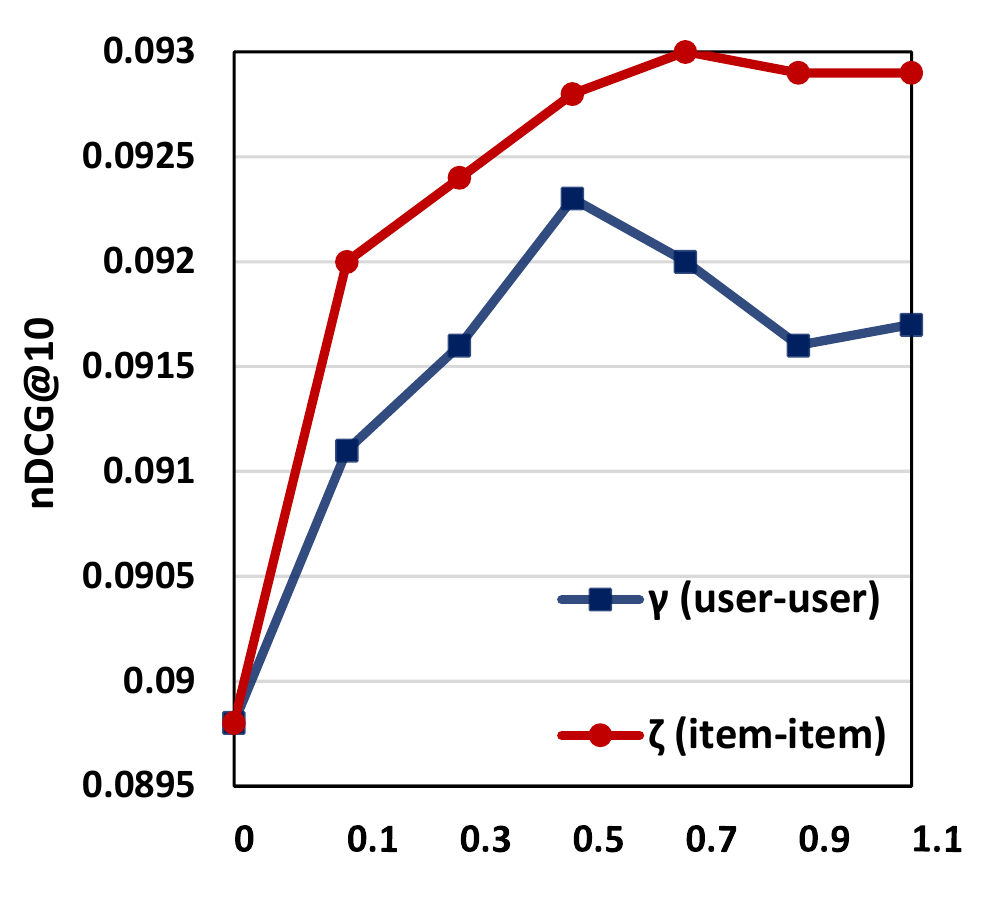} 
}  
\hspace{0cm}
\subfigure[Gowalla. ] {  
\includegraphics[width=0.45\columnwidth]{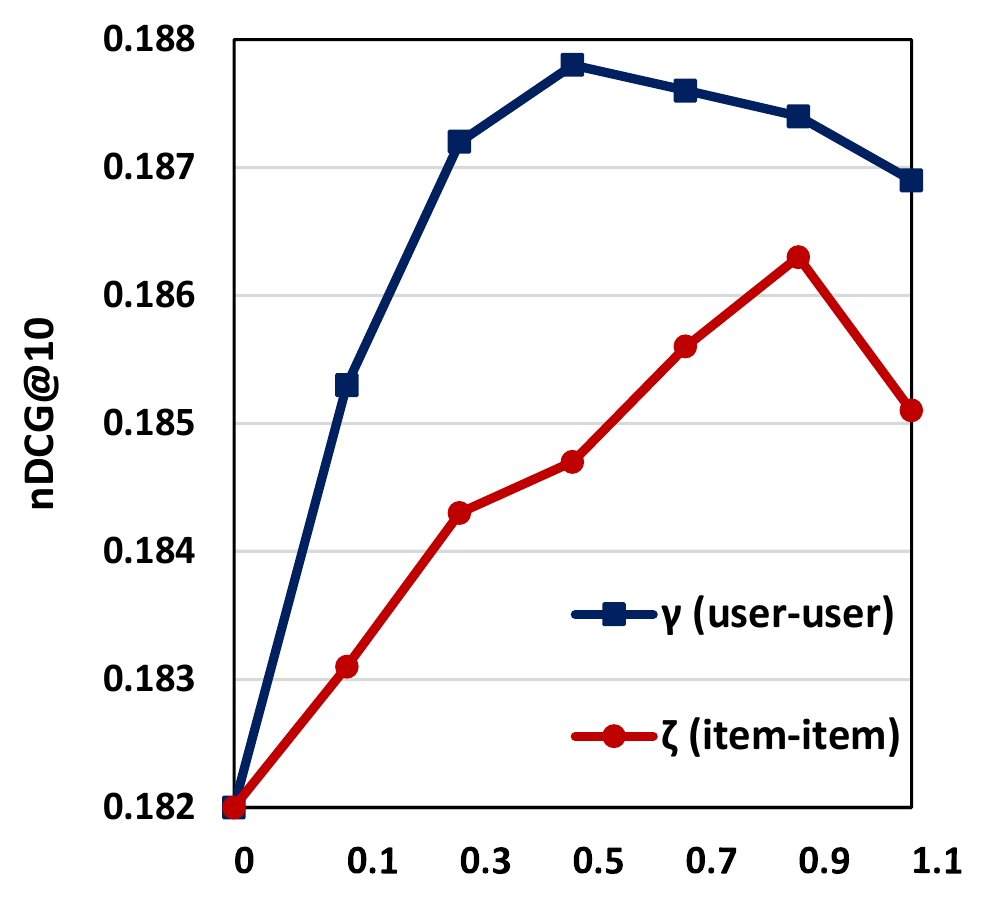} 
}
\vspace{-0.4cm}
\caption{Effect of $\gamma$ and $\zeta$.}  
\label{homo_relations}
\end{figure}

The direct comparison between SVD-GCN-B and SVD-GCN-U, SVD-GCN-I, and SVD-GCN-M demonstrates the positive effect of homogeneous relations. Furthermore, Figure \ref{homo_relations} shows how different $\gamma$ and $\zeta$ affect the accuracy, where the accuracy increases first then drops as constantly increasing the value of $\gamma$ and $\zeta$. The best accuracy is achieved at $\gamma=0.5$, while the optimal $\zeta$ (0.9 on Gowalla and 0.7 on Yelp) is larger than $\gamma$. One reasonable explanation is that item-item relations are usually sparser (0.21\% on Gowalla and 0.33\% on Yelp) than user-user relations (0.41\% on Gowalla and 0.48\% on Yelp).

\subsubsection{Do We Need Feature Transformation?}
By comparing SVD-GCN-S and SVD-GCN-B, we can see $\mathbf{W}$ results in worse accuracy on Gowalla and Yelp and only a slight improvement on ML-1M, which shows that feature transformation does not help much learn user-item interactions. On the other hand, we can identify the positive effect of $\mathbf{W}$ when incorporating user-user and item-item relations, which leads to improvement compared with SVD-GCN-B. We speculate that the ineffectiveness of feature transformation is related to the data density, where the intrinsic characteristic of sparse data such as user-item interactions is difficult to learn, while user-user and item-item relations are much denser thus is easier to learn. Overall, SVD-GCN can achieve superior accuracy without any model training, implying that the key design making GCN effective for recommendation lies in a good low-rank representation.

\subsubsection{Effect of Renormalization Trick}   
\begin{figure} \centering 
\subfigure[How $K$ and accuracy (nDCG@10) vary on $\alpha$ on SVD-GCN-B.] {  
\includegraphics[width=0.45\columnwidth]{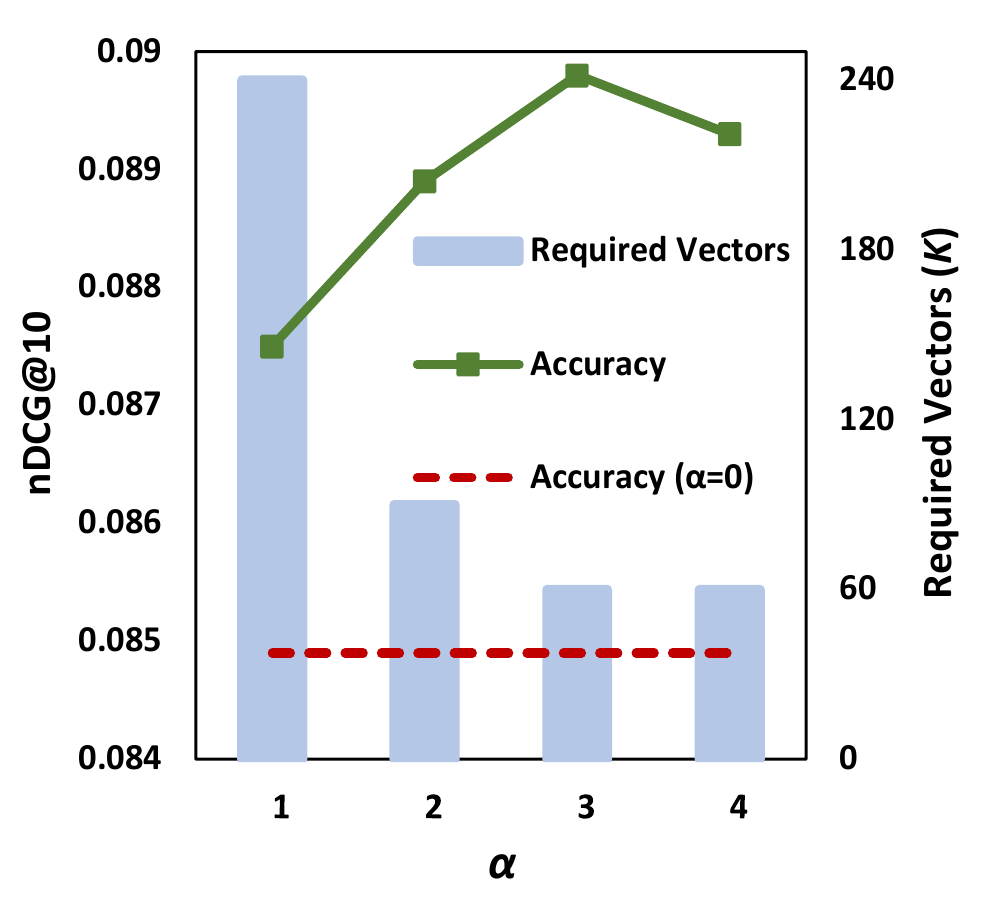} 
}  
\hspace{0cm}
\subfigure[Distribution of top 100 singular values with varying $\alpha$. ] {  
\includegraphics[width=0.45\columnwidth]{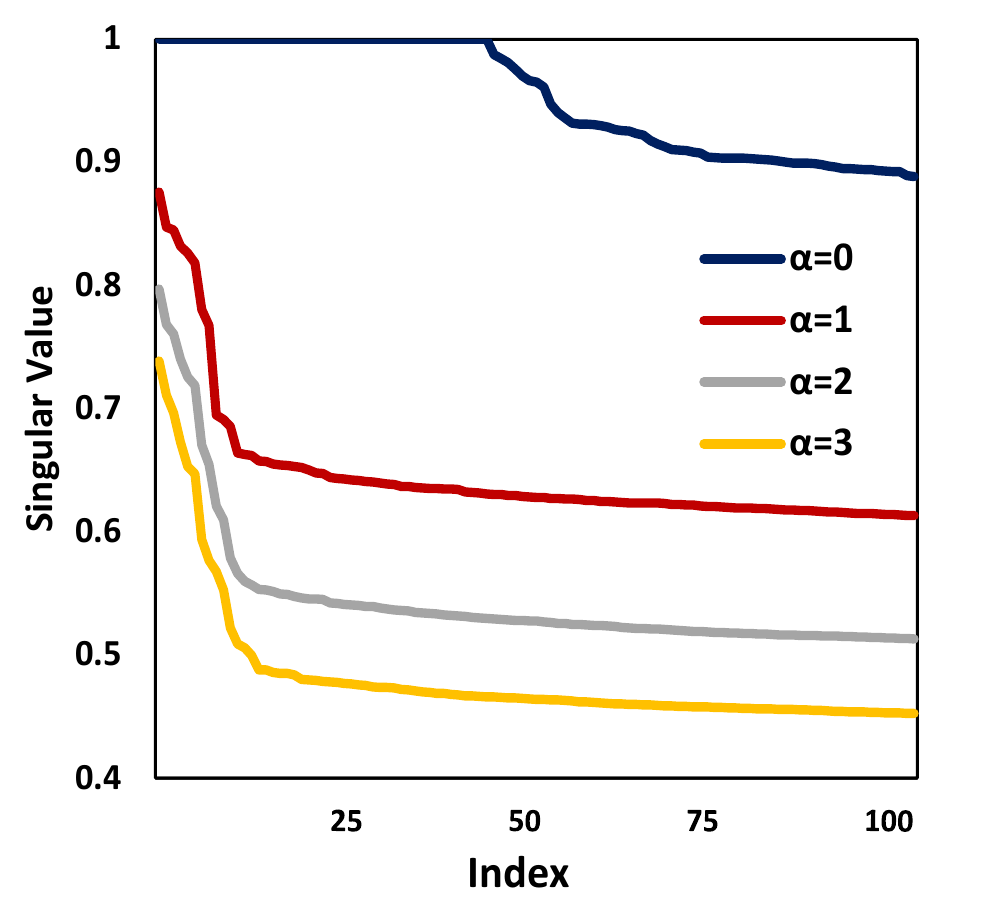} 
}
\vspace{-0.4cm}
\caption{Effect of renormalization trick on Yelp.}  
\label{renorm}
\end{figure}

We have two observations from Figure \ref{renorm} (a): as increasing $\alpha$ (i.e., shrinking the singular value gap), (1) the accuracy increases first then drops, reaches the best at $\alpha=3$; (2) the model tends to require fewer singular vectors. In Figure \ref{renorm} (b), as increasing $\alpha$, (1) the maximum singular value becomes smaller, which is consistent with Theorem 3; (2) singular values drops more quickly, which explains why fewer singular vectors are required. For instance, the model with $\alpha=0$ has more large singular values which contribute significantly to the interactions compared with the model with $\alpha>0$, thus more singular vectors are required; while the important large singular values are fewer as increasing $\alpha$. In other words, the important information is concentrated in fewer top singular values when we constantly increase $\alpha$. Surprisingly, we have the same observation on other datasets. Theoretical analysis on this interesting phenomenon is beyond the scope of this work, we leave it for future work. 

\subsubsection{Effect of $\beta$}
\begin{figure} \centering 
\subfigure[ML-1M.] {  
\includegraphics[width=0.45\columnwidth]{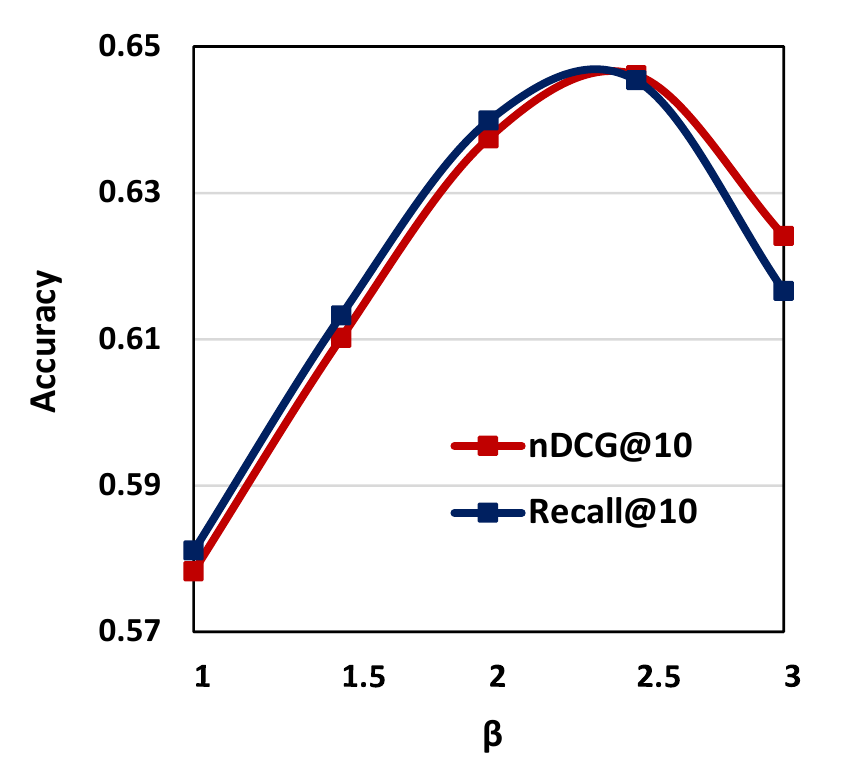} 
}  
\hspace{0cm}
\subfigure[Gowalla. ] {  
\includegraphics[width=0.45\columnwidth]{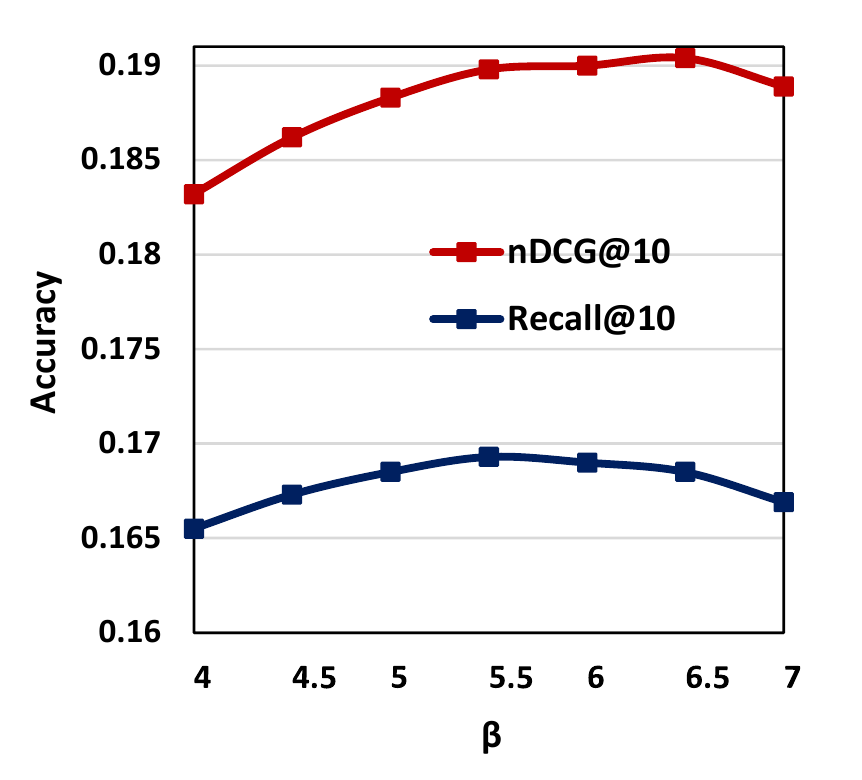} 
}
\vspace{-0.4cm}
\caption{Effect of $\beta$ on SVD-GCN-S.}  
\label{effect_beta}
\end{figure}

Figure \ref{effect_beta} shows the accuracy with varying $\beta$. The accuracy first increases as increasing $\beta$, then starts dropping after reaching the best performance at $\beta=2.5$ on ML-1M, $\beta=6.0$ on Gwoalla; there is a similar trend on Yelp that the best accuracy is achieved at $\beta=4.0$. We observe that $\beta$ tends to be larger on sparser data, implying that the large singular values are more important on the sparser data. We speculate that there is less useful information on sparser datasets, thus the small singular values contain more noise and should be depressed more than denser datasets.

\subsubsection{The Choice of Weighting Function}
We show the accuracy of SVD-GCN-S with different weighting functions in Table \ref{weight_func}. For dynamic designs, we use a neural network to attempt to model the importance of singular vectors with singular values as the input, while it underperforms most static designs, showing that the dynamic design is not suitable for the weighting function. For static designs, following the previous analysis in Section 3.2, we list some properties that matter to accuracy: (from left to right) if the function (1) is increasing, (2) has positive taylor coefficients, (3) is infinitely differentiable, and evaluate some functions, where the setting of $\beta$ is based on the best accuracy of each function. We can see the importance of the three properties is (1)$\gg$(2)$>$(3). (1) implies that the larger singular values are assigned higher weights, which is important according to the previous analysis; (2) and (3) suggest if the model can capture neighborhood from any-hops with positive contributions. Overall, the importance of the three properties is (1)$\gg$(2)$>$(3), and the functions satisfying all three properties perform the best.

\begin{table}[]
\caption{Accuracy of different weighting functions on Yelp.}
\scalebox{0.8}{
\begin{tabular}{c|c|c|ccc}
\hline
\multirow{3}{*}{Design} & \multirow{3}{*}{Function}      & \multirow{3}{*}{nDCG@10} & \multicolumn{3}{c}{Property}         \\ \cline{4-6} 
&                                &                          & (1) & (2)  & (3)   \\
                        &                                &                          & Increasing & Pos Coef.  & Infinite   \\ \hline
\multirow{4}{*}{Static} & $\log(\beta\sigma_k)$          & 0.0882                   & \checkmark & $\times$   & \checkmark \\
                        & $\sum_l^L\sigma_k^l$                  & 0.0899                   & \checkmark & \checkmark & $\times$   \\
                        & $\frac{1}{1-\beta\sigma_k}$   & 0.0919                   & \checkmark & \checkmark & \checkmark \\
                        & $e^{\beta\sigma_k} (\beta>0)$ & 0.0919                   & \checkmark & \checkmark & \checkmark \\
                        & $e^{\beta\sigma_k} (\beta<0)$ & 0.0828                    & $\times$   & $\times$   & \checkmark \\ \hline
Dynamic                 & Neural Network                 & 0.0850                   & \multicolumn{3}{c}{}                 \\ \hline
\end{tabular}}
\label{weight_func}
\end{table}

\section{Related Work}
Collaborative Filtering (CF) is an extensively used technique in modern recommender systems. Early memory-based CF methods \cite{sarwar2001item} predict user preference by computing the similarity between users or items. Later on, model-based methods become prevalent \cite{koren2009matrix} which characterizes users and items as latent vectors and calculate their dot products to predict the unobserved ratings. Subsequent works focus on modeling complex user-item interactions with advanced algorithms, such as neural network \cite{he2017neural,xue2017deep}, attention mechanism \cite{kang2018self}, transformer \cite{sun2019bert4rec}, and so on. Behind the learning in Euclidean space, some methods \cite{vinh2020hyperml} explore the potential of learning in non-Euclidean space. On another line, auxiliary information such as social relations \cite{ma2008sorec}, review data \cite{bao2014topicmf}, temporal information \cite{kang2018self} etc. is also well incorporated to obtain a better understanding of user preference.\par
   
The data sparsity issue on recommendation datasets limits the aforementioned traditional CF methods. The development of GCNs helps alleviate this issue by incorporating higher-order neighborhood to facilitate user/item representations, and thus much effort has been made to adapt GCNs to recommendation. Early work such as GC-MC \cite{berg2017graph} accumulates messages from different neighbors based on the rating for explicit feedbacks; SpectralCF \cite{zheng2018spectral} adapts the original graph convolution to CF with implicit feedbacks; NGCF \cite{wang2019neural} improves based on vanilla GCN \cite{kipf2017semi} by additionally encoding the interactions via an element-wise multiplication. To improve the scalability on large-scale datasets, Ying et al. \cite{ying2018graph} defines a flexible graph convolution on spatial domain without passing messages with adjacency matrix. By showing the redundancy of feature transformation and non-linear activation function, LightGCN \cite{he2020lightgcn} only keeps neighborhood aggregation for recommendation. Recent works fuse other research topics into GCNs, such as contrastive learning \cite{wu2020self,zou2022multi}, learning in hyperbolic space \cite{sun2021hgcf}, negative sampling \cite{huang2021mixgcf}, graph signal processing \cite{peng2022less}, etc. and achieves further success.\par     

Despite the superior performance that the aforementioned GCN-based methods have achieved, the computational cost of GCNs is much larger than traditional CF methods, making them unscalable on large-scale datasets. Although some works \cite{chen2020revisiting,he2020lightgcn} reduces the cost to some extent by removing feature transformation and non-linear activation functions, while the complexity mainly comes from the neighborhood aggregation, which is implemented by multiplying by an adjacency matrix. One recent work UltraGCN \cite{mao2021ultragcn} further simplifies GCNs by replacing the neighborhood aggregation with a weighted MF, where the weight is obtained from a single-layer LightGCN, which significantly reduces the complexity. However, such a simplification degrades the power of GCNs as it can only capture the first-order neighborhood, and the experimental results also show its ineffectiveness under extreme sparsity. On the other hand, our proposed SVD-GCN is based on comprehensive theoretical and empirical analysis on LightGCN with any layers, whose superiority and effectiveness have been demonstrated through extensive experimental results.

\section{Conclusion}      
In this work, we proposed a simplified and scalable GCN learning paradigm for CF. We first investigated what design makes GCN effective. Particularly, by further simplifying LightGCN, we showed that stacking graph convolution layers is to learn a low-rank representation by emphasizing (suppressing) more components with larger (smaller) singular values. Based on the close connection between GCN-based and low-rank methods, we proposed a simplified GCN formulation by replacing neighborhood aggregation with a truncated SVD, which only exploits $K$-largest singular values and vectors for recommendation. To alleviate over-smoothing issue, we proposed a renormalization trick to adjust the singular value gap, resulting in significant improvement. Extensive experimental results demonstrated the training efficiency and effectiveness of our propose methods.\par

We leave two questions for future work. Firstly, since SVD-GCN-S already achieves superior performance and feature transformation only shows positive effect learning user-user and item-item relations, we aim to incorporate user-user and item-item relations without introducing any model parameters (i.e., we improve based on SVD-GCN-S). In addition, we attempt to explain the phenomenon in Section 4.3.3, that why shrinking the singular value gap causes singular values to drop more quickly, thereby making important information to be concentrated in fewer singular vectors.

\section{Proofs}
\subsection{Proofs of Theorem 1}
\begin{proof}
Following SVD, we know any two singular vectors are orthonormal (i.e., $\mathbf{P}\mathbf{P}^T=\mathbf{I}$ and $\mathbf{Q}\mathbf{Q}^T=\mathbf{I}$), thus it is easy to derive the following equations:
\begin{equation}
\begin{aligned}
&\mathbf{\tilde{R}}\mathbf{\tilde{R}}^T=\mathbf{P}diag\left(\sigma_k^{2}\right)\mathbf{P}^T,\\
&\mathbf{\tilde{R}}^T\mathbf{\tilde{R}}=\mathbf{Q}diag\left(\sigma_k^{2}\right)\mathbf{Q}^T.
\end{aligned}
\label{proof_1}
\end{equation}
By repeating the above Equations $l$ times, we obtain Equation (\ref{thorem_11}).\par

For simplicity, we let $\mathbf{R}'=\mathbf{\tilde{R}}\left(\mathbf{\tilde{R}}^T\mathbf{\tilde{R}}\right)^{\frac{l\mbox{-}1}{2}}$, and $\mathbf{R}'^T=\mathbf{R}^T\left(\mathbf{\tilde{R}}\mathbf{\tilde{R}}^T\right)^{\frac{l\mbox{-}1}{2}}$. We let $\mathbf{P}'$, $\mathbf{Q}'$ and $\sigma'_k$ denote the stacked left singular vectors, right singular vectors and singular value for $\mathbf{R}'$, respectively. Following Equation (\ref{proof_1}), we can derive the following equations:
\begin{equation}
\begin{aligned}
&\mathbf{R}'\mathbf{R}'^T=\left(\mathbf{\tilde{R}}\mathbf{\tilde{R}}^T\right)^l=\mathbf{P}diag\left(\sigma_k^{2l}\right)\mathbf{P}^T=\mathbf{P}'diag\left(\sigma_k'^{2}\right)\mathbf{P}'^T,\\
&\mathbf{R}'^T\mathbf{R}'=\left(\mathbf{\tilde{R}}^T\mathbf{\tilde{R}}\right)^l=\mathbf{Q}diag\left(\sigma_k^{2l}\right)\mathbf{Q}^T=\mathbf{Q}'diag\left(\sigma_k'^{2}\right)\mathbf{Q}'^T.
\end{aligned}
\label{proof_2}
\end{equation}
It is easy to observe that $\mathbf{P}'=\mathbf{P}$, $\mathbf{Q}'=\mathbf{Q}$ and $\sigma'_k=\sigma_k^l$. Then, according to SVD, we derive Equation (\ref{thorem_12}). 

\end{proof}

\subsection{Proofs of Theorem 2 and 3}
\begin{proof}
We first introduce Rayleigh quotients \cite{spielman2012spectral}:
\begin{equation}
\lambda_{{\rm min}}\leq\mathbf{x}^T\mathbf{\tilde{A}}\mathbf{x}\leq\lambda_{{\rm max}} \quad s.t. \left|\mathbf{x}\right|=1,
\end{equation}
where $\lambda_{{\rm min}}$ and $\lambda_{{\rm max}}$ are the minimum and maximum eigenvalues of $\mathbf{\tilde{A}}$, respectively. Then, we can show $\lambda_{{\rm max}}=1$:
\begin{equation}
1-\mathbf{x}^T\mathbf{\tilde{A}}\mathbf{x}=\mathbf{x}^T\mathbf{x}-\mathbf{x}^T\mathbf{\tilde{A}}\mathbf{x}=\sum_{(u,i)\in\mathcal{E}}\left(\frac{x_u}{\sqrt{d_u}}-\frac{x_i}{\sqrt{d_i}} \right)^2\geq0.
\end{equation} 
In the meanwhile, we have the following observation:
\begin{equation}
\mathbf{\tilde{A}}
\begin{bmatrix}
 \mathbf{p}_k\\ 
\mathbf{q}_k
\end{bmatrix}
=
\begin{bmatrix}
\mathbf{\tilde{R}}\mathbf{q}_k\\ 
\mathbf{\tilde{R}}^T\mathbf{p}_k
\end{bmatrix}
=\sigma_k
\begin{bmatrix}
 \mathbf{p}_k\\ 
\mathbf{q}_k
\end{bmatrix},
\label{proof_3}
\end{equation} 
which implies that $\sigma_k \in \{\lambda_{{\rm min}},\cdots, \lambda_{{\rm max}}\}\leq1$ with $[\mathbf{p}_k, \mathbf{q}_k]^T$ as the eigenvector. By observing the eigenvector of $\lambda_{{\rm max}}$, if $\lambda_{{\rm max}}$ is also a singular value, we have: $\mathbf{p}_k=\sqrt{\mathbf{D}_U}\mathbf{1}$ and $\mathbf{q}_k=\sqrt{\mathbf{D}_I}\mathbf{1}$ where $\mathbf{1}$ is a vector with all 1 elements. It is easy to verify the solution satisfies SVD: $\mathbf{\tilde{R}}\mathbf{q}_k=\mathbf{p}_k$, thus $\sigma_{{\rm max}}=1$.\par

Given $\dot{\mathbf{R}}$, we can define the corresponding adjacency matrix $\dot{\mathbf{A}}$. Since the relation in Equation (\ref{proof_3}) still holds between $\dot{\mathbf{R}}$ and $\dot{\mathbf{A}}$, we only need to prove $\dot{\lambda}_{\rm{max}}\leq \frac{d_{{\rm max}}}{d_{{\rm max}}+\alpha}$. 
\begin{equation}
\begin{aligned}
\mathbf{x}^T\dot{\mathbf{A}}\mathbf{x}&=\sum_{(u,i)\in\mathcal{E}}\frac{2x_ux_i}{\sqrt{d_u+\alpha}\sqrt{d_i+\alpha}}\leq\sum_{u\in\mathcal{V}}\frac{d_u}{d_u+\alpha}x_u^2,\\
&=1-\sum_{u\in\mathcal{V}}\frac{\alpha}{d_u+\alpha}x_u^2\leq 1-\frac{\alpha}{d_{{\rm max}}+\alpha}=\frac{d_{{\rm max}}}{d_{{\rm max}}+\alpha}.
\end{aligned}
\label{proof_4}
\end{equation}
$=$ holds when $\alpha=0$. When $\alpha>0$, $\dot{\lambda}_{\rm{max}}<\frac{d_{{\rm max}}}{d_{{\rm max}}+\alpha}$, since $\mathbf{x}$ takes different values at $\sum_{(u,i)\in\mathcal{E}}\frac{2x_ux_i}{\sqrt{d_u+\alpha}\sqrt{d_i+\alpha}}=\sum_{u\in\mathcal{V}}\frac{d_u}{d_u+\alpha}x_u^2$ and $1-\sum_{u\in\mathcal{V}}\frac{\alpha}{d_u+\alpha}x_u^2= 1-\frac{\alpha}{d_{{\rm max}}+\alpha}$.

\end{proof}

\bibliographystyle{ACM-Reference-Format}
\bibliography{sample-base}
\end{document}